%% file: main.tex
\documentclass[sigconf,authorversion]{acmart}
\input{ethos}

\AtBeginDocument{%
  }
\setcopyright{acmlicensed}
\copyrightyear{2026}
\acmYear{2026}
\acmDOI{XXXXXXX.XXXXXXX}
\acmISBN{978-1-4503-XXXX-X/2026/06}
\acmSubmissionID{2321}

\begin{document}
\title{\sys{}: A Practical Framework for Privacy-Preserving \\ Device Logs}

\author{Sanket Goutam}
\affiliation{%
  \institution{Stony Brook University}
  \city{}
  \country{}
}

\author{Hunter Kippen}
\affiliation{
  \institution{Samsung Research America}
  \city{}
  \country{}
}

\author{Mike Grace}
\affiliation{
  \institution{Unaffiliated}
  \city{}
  \country{}
}

\author{Amir Rahmati}
\affiliation{
  \institution{Stony Brook University}
  \city{}
  \country{}
}

\input{abstract}



\maketitle

\input{chapters/introduction}
\input{chapters/background}

\input{chapters/motivation}
\input{chapters/threat-model}
\input{chapters/design}

\input{chapters/evaluation}

\input{chapters/security_analysis}
\input{chapters/discussion}

\input{chapters/related_work}
\input{chapters/conclusion}

\section{Generative AI Usage}
We used ChatGPT, Gemini, and Claude to assist in code prototyping and to refine the manuscript's prose. All AI-assisted code was manually verified via compilation and testing. The authors authored the original draft of this paper; LLMs were used only for grammatical and quality improvements. Every AI-generated suggestion was reviewed by the authors to ensure technical accuracy and the absence of hallucinations.

\bibliographystyle{ACM-Reference-Format}
\bibliography{references}

\appendix
\input{chapters/implementation}

\end{document}

%% file: ethos.tex
\usepackage{multirow}
\usepackage{xspace}
\usepackage{subcaption} 
\usepackage{pifont} 
\usepackage{amsthm} 
\usepackage[ruled,vlined,linesnumbered]{algorithm2e}
\usepackage{listings}
\usepackage{tikz}

\let\OLDthebibliography\thebibliography
\renewcommand\thebibliography[1]{
  \OLDthebibliography{#1}
  \setlength{\parskip}{0pt}
  \setlength{\itemsep}{0pt plus 0.3ex}
}

\SetAlgoLined
\SetKwInOut{Input}{Input}
\SetKwInOut{Output}{Output}
\SetKwComment{Comment}{// }{}
\SetCommentSty{textit}
\DontPrintSemicolon

\newcommand{\KDFtext}{\ensuremath{\mathsf{KDF}}}
\newcommand{\AEAD}{\ensuremath{\mathsf{AE}}}
\newcommand{\Enc}{\ensuremath{\mathsf{Enc}}}
\newcommand{\Dec}{\ensuremath{\mathsf{Dec}}}
\newcommand{\ECDHtext}{\ensuremath{\mathsf{ECDH}}}
\newcommand{\HMAC}{\ensuremath{\mathsf{HMAC}}}

\usepackage[normalem]{ulem} 

\usepackage{enumitem} 
\setlist{nosep} 
\setlist{noitemsep,topsep=2pt,parsep=0pt,partopsep=0pt} 
\setlist[itemize]{leftmargin=*} 

\usepackage{wrapfig}
\usepackage{rotating} 

\usepackage{booktabs} 

\usepackage{array} 
\newcommand{\PreserveBackslash}[1]{\let\temp=\\#1\let\\=\temp}
\newcolumntype{C}[1]{>{\PreserveBackslash\centering}p{#1}}
\newcolumntype{R}[1]{>{\PreserveBackslash\raggedleft}p{#1}}
\newcolumntype{L}[1]{>{\PreserveBackslash\raggedright}p{#1}}

\newcommand{\sys}[0]{Proteus\xspace}

\newif\ifsubmit
\submitfalse

\ifsubmit
    \newcommand{\amir}[1]{}
    
    \newcommand{\todo}[1]{}
    \newcommand{\sanket}[1]{}
    \newcommand{\hunter}[1]{}
\else
    
    \newcommand{\amir}[1]{\textcolor{blue}{Amir: #1}}
    
    \newcommand{\todo}[1]{\textcolor{red}{TODO: #1}}
    \newcommand{\sanket}[1]{\textcolor{cyan}{Sanket: #1}}
    \newcommand{\hunter}[1]{\textcolor{cyan}{Hunter: #1}}
\fi

\newcommand{\paratitle}[1]{\noindent\textbf{\textit{#1.}}\xspace}
\newcommand{\xref}[1]{\S\ref{#1}}

\newcommand{\ie}[0]{\emph{i.e.,}\xspace}

\newcommand{\eg}[0]{\emph{e.g.,}\xspace}


%% file: abstract.tex
\begin{abstract}
Device logs are essential for forensic investigations, enterprise monitoring, and fraud detection; however, they often leak personally identifiable information (PII) when exported for third-party analysis. Existing approaches either fail to minimize PII exposure across all stages of log collection and analysis or sacrifice data fidelity, resulting in less effective analysis. We present \sys, a privacy-preserving device logging framework that enables forensic analysis without disclosing plaintext PII or compromising fidelity, even when facing adversaries with access to multiple snapshots of the log files. To achieve this, \sys proposes a two-layer scheme that employs keyed-hash pseudonymization of PII fields and time-rotating encryption with ratcheted ephemeral keys to prevent multi-snapshot correlation. For controlled sharing, clients export ratchet states that grant time-bounded access, permitting decryption of pseudonymized tokens that enable linkage and timeline reconstruction without exposing the underlying PII. Subsequent ratchet rotations ensure forward secrecy, while DICE-based attestation authenticates device provenance. We implement \sys as a transparent extension to Android’s logcat and evaluate it across three generations of hardware. Our results demonstrate a median latency of $0.2$ ms per message and an average per-PII-field size overhead of only $97.1$ bytes.

\end{abstract}

%% file: chapters/introduction.tex
\section{Introduction}
\label{sec:intro}

Security and operational analytics have long relied on collecting device logs, but the source of these logs is undergoing a fundamental shift. While telemetry was historically sourced from corporate-controlled workstations within a clear security perimeter, it is now increasingly gathered from a diverse ecosystem of user-centric devices. This trend is prominent in enterprise settings through Bring-Your-Own-Device (BYOD) policies, which integrate personal smartphones into corporate monitoring. Telemetry collection has now extended into the consumer space, where services collect logs from smart TVs, IoT hubs, medical wearables, and streaming clients for fraud detection and abuse prevention~\cite{Karagiannis2023,esmaeilzadeh2022,microsoft_intune}. The expansion of log collection to personally owned devices creates a privacy dilemma. Unlike their corporate counterparts, these endpoints are deeply embedded in users' private lives, capturing sensitive data from communications, location history, and health applications. Consequently, granular logs essential for security and analytics inevitably expose users' personally identifiable information (PII) when exfiltrated, pitting the need for observability against users' right to privacy and often violating regulatory requirements~\cite{eu_gdpr,ccpa,Lyons2023}.

This privacy dilemma persists because existing data protection techniques are fundamentally mismatched with the mobile endpoint threat model. Solutions like post-hoc redaction~\cite{microsoft_presidio,mainetti2025detecting}, client-side taint tracking~\cite{Enck2010TaintDroid,Arzt2014FlowDroid,hu2021samldroid,yang2022fsaflow}, and encrypted auditing~\cite{chaulagain2024fa,hugenroth2024sloth} were designed for traditional enterprise environments with clear trust boundaries. As we detail in \xref{sec:motivation}, they fail in the mobile context: they either expose sensitive data during collection, compromise the event-level fidelity required for forensics, or encounter critical deployment friction. Traditional data protection approaches are ill-suited to a world of user-owned devices, with continuous data exfiltration to honest-but-curious cloud platforms.

In this paper, we first formalize the requirements for privacy-preserving logging under a mobile endpoint threat model (\xref{sec:threatmodel}). We consider an adversary who can capture multiple log snapshots over time (a multi-snapshot on-device observer) or receive exported logs and perform large-scale correlation (an honest-but-curious server). These assumptions reflect real-world threat vectors in which a malicious app is installed on a device, a privileged user has access to device logs, or an adversary gains access to cloud storage through a breach. 

To resolve this tension, we present \sys{}, the first logging framework that provides in-situ privacy protection without compromising forensic utility. Our design is built on a key insight: effective forensic analysis requires \textbf{correlation} (linking related events), not the \textbf{disclosure} of raw PII values. \sys{} materializes this insight through a novel, two-layer cryptographic protocol applied during log generation. \sys{} operates on any log fields designated as sensitive. At log emission, these fields are first pseudonymized with a keyed hash, creating stable tokens that enable correlation. Second, it encrypts these tokens with daily-rotated keys derived from a hardware-rooted hierarchical ratchet, defending against correlation attacks by multi-snapshot adversaries. When analysis is needed, a controlled sharing protocol grants time-bounded access to decrypt the pseudonyms without exposing the underlying PII. This entire process is anchored in DICE attestation, which cryptographically binds logs to an attested device state~\cite{tao2021dice}.

We demonstrate that \sys provides robust protection through both theoretical analysis and deployment. We implemented and evaluated \sys{} as both an end-to-end prototype and an Android library on three generations of hardware (details in Appendix~\ref{sec:implementation}). We evaluate our deployment against the 30.3 million log entries in the LogHub dataset~\cite{zhu2023loghub}. Our results show that \sys{}'s in-situ protection is highly efficient, adding only 2.41\% in storage overhead and a median latency of 0.2~ms per message on real devices. A formal game-theoretic security analysis further confirms that our design achieves its goals of confidentiality and forward secrecy.

\paratitle{Contributions} In summary, this paper makes the following contributions:
\begin{itemize}
    \item \textbf{First in-situ privacy-preserving framework for mobile logging}: \sys{} protects sensitive data at the point of emission, preventing plaintext PII from ever leaving the device while preserving full forensic utility.
    \item \textbf{Formal threat model for mobile endpoint forensics}: We define the first threat model and security requirements for privacy-preserving logging systems centered around mobile devices with multi-snapshot adversaries and honest-but-curious cloud platforms.
    \item \textbf{Practical implementation and evaluation}: Android library deployed across three device generations with demonstrated practical overhead (median 0.2~ms latency, 2.41\% storage) suitable for production deployment.
    \item \textbf{Provable security guarantees}: Game-theoretic proof that \sys{}'s hierarchical ratcheting provides security properties analogous to Signal's double ratchet~\cite{cohn2017signal,perrin2016doubleratchet}, which has been formally verified to have confidentiality and forward secrecy guarantees.
\end{itemize}

%% file: chapters/background.tex
\section{Background}
\label{sec:background}

Enterprise endpoint-monitoring solutions are expanding their coverage beyond traditional desktops and servers to mobile devices and user-personal endpoints. Unified Endpoint Management (UEM) platforms today support everything from basic mobile device management to comprehensive endpoint telemetry, policy enforcement, and security monitoring across smartphones and tablets in BYOD settings~\cite{microsoft_intune,vmware_workspace_one,ibm_maas360,citrix_endpoint_management,samsung2023attestation}. This shift reflects a fundamental change: organizations now treat mobile endpoints as integral to their security posture rather than peripheral assets.

Mobile devices, unlike their traditional counterparts, pose unique challenges with respect to data extraction policies. Unlike enterprise-owned desktops and servers, smartphones and tablets operate in deeply personal spaces and serve as repositories for sensitive user data. They contain private communications (messaging, email), precise location traces (GPS coordinates, visited venues), health information (fitness tracking, medical apps), and behavioral patterns that extend far beyond work-related activities. Yet modern UEM and SIEM (Security Information and Event Management) solutions require continuous log exfiltration from these devices to feed AI-driven analytics pipelines for anomaly detection, fraud prevention, and behavioral analysis~\cite{manageengine_siem_logtypes_2025}. This creates an inherent tension: the same granular logs that enable effective security monitoring inevitably expose PII when collected and analyzed by third-party platforms, even when the original purpose is purely security-focused.

Beyond enterprise BYOD scenarios, consumer-facing setups such as streaming platforms, smart TVs, and IoT hubs deploy similar endpoint data-collection pipelines for fraud detection and device-integrity monitoring. Modern SIEM solutions collect device logs to enable unified visibility across device, user, and network events~\cite{manageengine_siem_logtypes_2025,esmaeilzadeh2022}. These logs capture fine-grained events (timestamps, process identifiers, network interactions, sensor readings, and runtime states) that allow security analysts to reconstruct timelines, build attack graphs, perform provenance analysis, and trace incident root causes~\cite{inam2023sok,zeng2021watson,xu2022depcomm}. In Table~\ref{tab:endpoint-comparison} we contrast traditional workstation logging with modern mobile endpoint logging, highlighting the fundamental differences in data ownership, sensitivity, and collection models. Thus, the challenge for mobile endpoint integration into SIEM is threefold: ensuring event-level fidelity (\ie capturing what, when, and where for each event), managing privacy (limiting PII leakage), and handling scale (volume of mobile logs). Unlike enterprise setups, where corporate ownership permits broad data collection, consumer-grade mobile devices require carefully designed telemetry pipelines that respect user privacy and uphold data compliance standards while providing sufficient granularity for forensic analysis.

\begin{table*}[htbp]
    \centering
    \caption{Comparison of Traditional vs. Modern SIEM Workflows}
    \begin{tabular}{p{2.5cm}p{6.7cm}p{6.7cm}}
    \hline
    \textbf{Aspect} & \textbf{Traditional SIEM (Enterprise Systems)} & \textbf{Modern SIEM (Mobile / Personal Devices)} \\ 
    \hline
    
    \textbf{Data Ownership} & 
    Owned and managed by the enterprise. Logs originate from corporate-controlled infrastructure and hosts. Centralized IT policies dictate collection, retention, and access control. & 
    Data resides on user-owned mobile or IoT endpoints. Collection often depends on user consent or application permissions. Ownership is fragmented between user, OS vendor, and service provider. \\ 
    
    \textbf{Data Context for Analysis} & 
    Context primarily involves enterprise applications, network flows, and access control events. Analysis relies on structured corporate assets and identity systems (\eg Active Directory). & 
    Context includes diverse app ecosystems, sensors, connectivity states, and background system events. Analysis requires correlating heterogeneous data sources with dynamic user activity. \\ 
    
    \textbf{PII Sensitivity} & 
    Limited; logs mostly contain operational or authentication data tied to employee identifiers within enterprise boundaries. & 
    High; mobile logs often include location traces, user identifiers, contact information, and behavioral signals. Privacy-preserving collection is essential to mitigate the exposure of user information. \\ 
    
    \textbf{Collection Scope} & 
    Within a single enterprise trust boundary. Data aggregation occurs in a secured network perimeter managed by corporate SOCs (Security Operations Center). & 
    Crosses user trust boundaries: data may traverse from user space to cloud SIEM platforms or third-party analytics engines. Requires federated trust and data minimization guarantees. \\ 
    
    \textbf{Privacy Risks} & 
    Risks are limited to insider misuse or over-retention of audit data. & 
    Risk of re-identification, inference of private attributes, and regulatory non-compliance (\eg GDPR, CCPA) if logs are not anonymized or filtered. \\ 
    \hline
    
    \end{tabular}
    \label{tab:endpoint-comparison}
\end{table*}

The shift from traditional enterprise workstations to mobile and IoT endpoints fundamentally changes the privacy calculus for security monitoring. While enterprise-owned systems operate within clear trust boundaries and organizational policies, mobile devices cross into personal user space, where privacy expectations are high and regulatory protections apply. Current logging practices, however, were designed for the traditional model and fail to account for the heightened PII sensitivity and continuous exfiltration requirements of modern mobile SIEM workflows. This mismatch underscores the need for privacy-preserving logging approaches that operate effectively in mobile contexts.

%% file: chapters/motivation.tex
\section{Limitations of Current Approaches}
\label{sec:motivation}

Most existing mobile logging frameworks provide no protection for PII at the point of emission, relying instead on downstream controls. Current approaches primarily employ post-hoc sanitization~\cite{microsoft_presidio}, enterprise DLP (Data Loss Prevention)~\cite{microsoft_dlp_security101}, and cryptographic controls over stored audit data~\cite{chaulagain2024fa}. While these methods address specific risks, they rely on trust assumptions that fail in mobile contexts, leaving residual PII exposure on devices, during extraction, or in post-processing stages. We examine each approach and identify why it fails to meet the requirements for privacy-preserving mobile logging.

A dominant industrial practice is the application of automated PII detection and redaction tools such as Microsoft Presidio~\cite{microsoft_presidio}, which employ rule-based recognizers and statistical models to identify sensitive entities (\eg names, email addresses, device identifiers) in textual data and redact them before further processing. Enterprise DLP and UEM platforms~\cite{digitalguardian_byod,techtarget_uem,liu2018towards,citrix_endpoint_management} integrate similar detection mechanisms to enforce centralized inspection of data-in-transit and apply policy-based filters. Although effective within controlled enterprise pipelines, these approaches share a fundamental limitation: they embody a post-hoc privacy model. Raw logs are extracted from user devices and ingested by the service infrastructure prior to sanitization, thereby placing trust in the collection and staging servers. Even when protected in transit (\eg TLS), logs are commonly stored in cleartext within processing backends for indexing and analysis, extending broad access to service operators and conflicting with least-privilege principles~\cite{cwe359}. In large-scale deployments, these systems also face operational complexity, high false-positive rates, and inconsistent enforcement across diverse endpoints. While valuable for compliance and policy governance, post-hoc redaction and DLP frameworks fail to reconcile the asymmetry between device-side data generation and remote data control, leaving privacy enforcement external to the source.

Client-side analysis frameworks such as TaintDroid~\cite{Enck2010TaintDroid} and FlowDroid~\cite{Arzt2014FlowDroid} attempt to detect privacy violations before data leaves the device by tracking the flow of sensitive information from sources to sinks. Although effective as research prototypes, these systems incur measurable runtime overheads, lack comprehensive coverage of proprietary binaries and third-party SDKs, and struggle to remain compatible across modern Android versions. Moreover, static taint policies cannot dynamically adapt to evolving data types or complex logging paths, resulting in both false alarms and missed leaks. Consequently, these tools have seen little adoption in production-grade telemetry or forensic pipelines.

Differential privacy mechanisms, such as Google's RAPPOR~\cite{erlingsson2014rappor}, take a fundamentally different approach by applying client-side aggregation and randomized response to provide formal privacy guarantees. However, these techniques are designed for statistical analysis of large populations and inherently compromise per-event fidelity. Forensic investigations, fraud detection, and security incident response all require precise event reconstruction and timeline correlation, all of which are explicitly sacrificed by differential privacy. 

Cryptographic approaches to audit log protection have evolved from early integrity-focused designs~\cite{schneier1999secure} to more recent systems that support selective decryption. FA-SEAL~\cite{chaulagain2024fa}, for instance, encrypts audit logs symmetrically and enables selective decryption of relevant segments through clustering and segmentation. While this addresses the wholesale-decryption problem, it does not solve the PII exposure issue: any decrypted fragment may still contain embedded identifiers that reveal sensitive information. 

Collectively, these approaches reveal a persistent gap. Post hoc redaction and DLP frameworks can expose PII during collection and staging. Client-side taint tracking suffers from coverage gaps and deployment friction. Differential privacy sacrifices per-event fidelity. Encrypted audit systems either require wholesale decryption for analysis or impose prohibitive performance costs. No existing approach enforces in-situ privacy protection to prevent plaintext PII from leaving the device, while preserving the correlation, timeline reconstruction, and linkage analysis required by forensic investigations.

\paratitle{Design goals} To address these limitations, we outline the following design goals for an in-situ privacy-preserving logging framework:
\begin{enumerate}
    \item[G1] \textbf{In-situ protection}: Protect sensitive fields at the point of generation so that PII never exists in exportable plaintext on or off the device.
    \item[G2] \textbf{Utility preservation}: Preserve non-PII context and event structure to enable indexing, filtering, and timeline reconstruction.
    \item[G3] \textbf{Linkage preservation}: Enable correlation and timeline reconstruction analysis over privacy-preserving derivatives (\eg keyed hashes) of PII without disclosing original values.
    \item[G4] \textbf{Forward secrecy and device binding}: Defend against multi-snapshot adversaries who capture logs over time. Bind logs to attested device identity to prevent tampering and injection attacks.
    \item[G5] \textbf{Post compromise security}: Provide a mechanism to restore secrecy of future logs after compromise of device state.
    \item[G6] \textbf{Deployability}: Integrate with production logging frameworks (\eg Android \texttt{logcat}) without specialized hardware or application modifications, imposing minimal overhead.
\end{enumerate}

\paratitle{Problem Statement} We seek to prevent PII leakage in device logs while preserving the forensic utility required for incident response, fraud detection, and security analysis. Specifically, logs must support correlation and timeline reconstruction without ever exposing plaintext PII, even to honest-but-curious servers or adversaries who capture multiple log snapshots over time.

The limitations outlined above stem from a common root cause: existing approaches were designed for traditional enterprise threat models in which trust boundaries are well defined, and devices are corporate-owned. They fail in the mobile context because they do not account for multi-snapshot adversaries who can observe logs over extended periods, honest-but-curious cloud platforms that perform AI-driven analytics, or the deeply personal nature of data on user-owned devices. To address this gap, we must first formalize the requirements for a privacy-preserving logging system within a threat model appropriate for mobile endpoints. We establish this foundation in \xref{sec:threatmodel}.

%% file: chapters/threat-model.tex
\section{Threat Model and System Formalization}
\label{sec:threatmodel}

In this section, we formalize the notion of a privacy-preserving logging system and define the threat model it must defend against. 

\paratitle{System Model} We consider endpoint logging pipelines in which consumer devices emit events via OS logging subsystems (\eg Android \texttt{logcat}). Log records may contain PII alongside non-PII context and are exported over authenticated channels to service-side storage and analytics. This establishes clear trust boundaries among the device, the transport, and the backend services. Throughout, we assume standard transport-layer protections (TLS/QUIC) and focus on exposures arising from log content at rest and during processing.

\paratitle{Adversary Model} The goal of the adversary is to recover plaintext PII, link identities across epochs, or infer sensitive attributes from logs. We model two complementary vantage points. First, a multi-snapshot on-device observer (a malicious app, privileged logger, or an attacker with repeated physical access) can obtain successive log snapshots via legitimate APIs or undetected extraction and attempt cross-time correlation, as observed in high-risk surveillance settings~\cite{nist800-92}. Second, a server-side adversary (honest-but-curious operators or breach adversaries) can access stored logs, associated metadata, previously shared decryption keys, and analytics-derived tokens across many devices and epochs, enabling linkage at scale. Neither adversary controls the OS kernel or device trust anchors.

\paratitle{Trust Assumptions} We assume the device OS and the logging modules execute correctly and are not compromised; cryptographic primitives and transport protections are implemented correctly with high-quality randomness; initial device identity, attestation, and key provisioning are trustworthy; and PII detection (\eg Regex) attains high recall for common sensitive types, leaving a residual risk from undetected fields that deployment policy must mitigate.

\paratitle{Security Objectives} Under this threat model, a solution must satisfy the design goals articulated in \xref{sec:motivation}. Specifically, it must ensure that PII remains confidential against both on-device observers (who capture multiple snapshots) and honest-but-curious servers (who receive exported logs), while preserving the forensic utility required for correlation and timeline reconstruction. Both forward secrecy and post-compromise security are essential: compromise of the current device state must not expose prior logs or allow for recovering the security of future logs. Finally, logs must be cryptographically bound to attested device identity to detect tampering and prevent injection attacks.

\paratitle{Out of Scope} Physical attacks, firmware modification, side-channel extraction, and denial-of-service are considered out of scope. Metadata privacy (timestamps, sizes) and traffic-analysis resistance (against side-channel attacks) are considered orthogonal to PII protection.

%% file: chapters/design.tex
\section{\sys{}}
\label{sec:design}

\sys{} is the first in-situ logging framework that prevents the disclosure of PII while preserving full forensic utility. Our design rests on a key insight: forensic analysis requires \emph{correlation}---creating linkage between related events---but not \emph{disclosure} of the actual PII values. This suggests that simple pseudonymization of PII fields could protect against most privacy attacks that involve harmful PII disclosure. However, simple pseudonymization is insufficient against the multi-snapshot adversary model of \xref{sec:threatmodel}. An adversary who captures logs over time can correlate pseudonymized tokens to infer linkage between user activity and events, revealing associations even without recovering the plaintext PII. For example, observing that \texttt{BASE64-HASH-EMAIL} appears in logs during office hours, alongside work-related activity, reveals behavioral patterns and enables profiling attacks.

A simpler approach would be to obfuscate all PII. Microsoft Presidio~\cite{microsoft_presidio}, the current state-of-the-art system, employs this strategy to manage PII-injected logs used in analytics tools. However, as Figure~\ref{fig:log-comparison} illustrates, this post-hoc redaction not only lacks protection at rest or during collection and analysis, but also sacrifices the forensic fidelity needed for correlation. \sys{} resolves this trade-off by operationalizing the key insight that analysis requires linkage, not recovery, through a two-layer protection protocol that enables user-authorized, controlled disclosure of data for forensic analysis without sacrificing utility for privacy.

\begin{figure}[t]
  \centering
  \begin{lstlisting}[language=bash,
            frame=shadowbox,
            breaklines=true,
            basicstyle=\ttfamily\scriptsize
            ]
  # Plaintext PII in logs (Status Quo)
  10-15 14:23:47.821  2341  2341 I AuthService: Login attempt for user alice@example.com from device IMEI:352099001761481 

  # Microsoft Presidio: Redaction destroys linkage
  10-15 14:23:47.821  2341  2341 I AuthService: Login attempt for user <EMAIL_ADDRESS> from device <IMEI>

  # Simple Pseudonymized hash
  10-15 14:23:47.821  2341  2341 I AuthService: Login attempt for user <PII type="EMAIL">BASE64-HASH-EMAIL</PII> from device <PII type="IMEI">BASE64-HASH-IMEI</PII>

  # Our System: Encrypted PII (on-device and in-transit)
  10-15 14:23:47.821  2341  2341 I AuthService: Login attempt for user <PII type="EMAIL">CIPHERTEXT-EMAIL</PII> from device <PII type="IMEI">CIPHERTEXT-IMEI</PII>
  \end{lstlisting}
  \caption{Simple pseudonymization preserves linkage but exposes correlation patterns to multi-snapshot adversaries. \sys{}'s two-layer protection scheme protects against unauthorized correlation while enabling controlled, user-authorized decryption for forensic analysis.}
  \label{fig:log-comparison}
\end{figure}

Consider four approaches to handling a typical Android log entry containing an email address and device identifier (Figure~\ref{fig:log-comparison}):

The status quo exposes plaintext PII to all parties. Microsoft Presidio's post-hoc redaction replaces PII with generic type labels (\texttt{<EMAIL\_ADDRESS>}, \texttt{<IMEI>}), preventing disclosure but destroying forensic utility. All users become indistinguishable, making it impossible to track repeated login attempts or correlate suspicious activity across sessions. Simple pseudonymization addresses both problems by replacing PII with stable keyed hashes (\texttt{BASE64-HASH-EMAIL}), enabling linkage while preventing plaintext recovery. However, this approach fails against multi-snapshot adversaries who can observe correlation patterns over time---\eg inferring that \texttt{BASE64-HASH-EMAIL} corresponds to a work email address based on temporal activity patterns.

With \sys{}, we introduce an additional layer of protection. At log emission, PII fields are first pseudonymized via keyed hashing, then encrypted with daily-rotated keys. The resulting \texttt{CIPHERTEXT-EMAIL} value is a base64-encoded ciphertext that prevents unauthorized correlation: even adversaries who capture multiple log snapshots cannot link events without decryption keys. This encrypted form is stored on-device and transmitted to servers, ensuring that plaintext PII never exists in an exportable form. When forensic analysis is required, servers receive user-authorized controlled access (via the protocol in \xref{sec:design:sharing:client}) to decrypt ciphertexts and recover the underlying pseudonymized hashes (\texttt{BASE64-HASH-EMAIL}). These stable, base64-encoded 16-byte hashes enable correlation of all events involving the same email address across epochs originating from the same device. Crucially, even the forensic server with decryption keys cannot reverse the keyed hash to obtain \texttt{alice@example.com}---analysts can determine \emph{whether} two events involve the same user, but not \emph{who} that user is. This two-layer protection scheme ensures confidentiality against unauthorized observers (via encryption) and unlinkability to real identities (via pseudonymization) while preserving full forensic utility through controlled access.

\begin{figure*}[!ht]
    \centering
    \includegraphics[width=\textwidth]{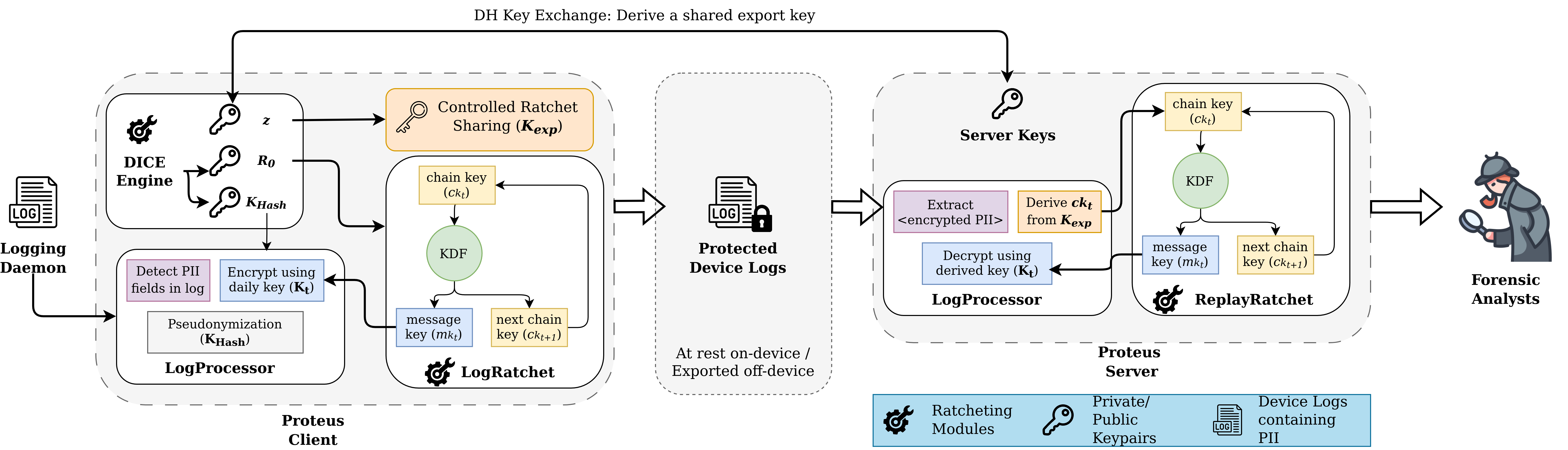}
    \caption{\sys{} is designed as a client-server system geared towards forensic data logging and collection, introducing an in-situ framework for sensitive data protection.}
    \label{fig:component-diagram}
\end{figure*}

We designed \sys{} to enable remote device log collection and analysis for forensic investigations. The end-to-end architecture is shown in Figure~\ref{fig:component-diagram}. We will now detail the cryptographic designs behind the client and server components.

\subsection{Client-Side Components}
\label{sec:design:client}

\paratitle{Key Derivation and Initialization}
\label{sec:design:init}
We choose DICE to be at the heart of our framework because it addresses three problems that privacy-preserving logging requires but that software-only approaches cannot reliably address. First, DICE yields a hardware-rooted CDI derived from a device-unique secret and measured software state. Keys derived from the CDI are, by construction, bound to both the physical device and its integrity measurements \cite{tao2021dice,tcg_dice_arch}. Second, this binding naturally supports device attestation: the measurements that define the CDI also produce evidence that a verifier can check before trusting exported material. Third, DICE has been standardized by the Trusted Computing Group (TCG) and has seen increased adoption across manufacturers as a foundation for device-bound identifiers and attestation, making it a pragmatic choice for large-scale deployment \cite{tcg_dice_arch}. In this model, the hardware's role is to securely produce the CDI based on software measurements. \sys{}, operating in user-space, then uses this CDI as the root of its key hierarchy. Our implementation models this interface by taking a simulated CDI as input.

\sys{} anchors all long-term secrets in the CDI so that they are usable only on a correctly measured device state. The client derives two keys from the CDI: (i) a per-system hash key $K_{\mathrm{hash}}$ used exclusively on-device for PII pseudonymization; and (ii) the root ratchet key $R_0$. $R_0$ is then used to initialize the symmetric ratchet, which generates daily encryption keys, thereby ensuring forward secrecy against multi-snapshot adversaries. Together, these realize \emph{attestation} (binding logs to attested device state), \emph{integrity} (measured-boot binding of derived keys), and \emph{authentication} (mutual authentication and export encryption during sharing), addressing design goals G1 (in-situ protection), G3 (linkage preservation), and G4 (forward secrecy and device binding). Algorithm~\ref{alg:client-init} details the initialization process. Notice that the derivation of $R_0$ is identical to the $\mathrm{RatchetInitAlice}$ procedure in \S{3.3} of the original description of the double ratchet algorithm \cite{perrin2016doubleratchet}, where the $\mathsf{CDI}$ takes the place of $\mathrm{SK}$. Because the CDI is withheld from the server, it cannot derive $R_0$ (and the subsequent chain keys) without additional interaction with the client.

\begin{algorithm}[t]
\caption{Client initialization}
\label{alg:client-init}
\Input{CDI, Server public key $B$}
\Output{Initial keys and state}
\Comment{Derive root keys from hardware-rooted CDI and Server pk}
$K_{\mathrm{hash}} \gets \KDFtext(\mathsf{CDI}, \text{``hmac-key''}, 32)$\;
$\mathrm{nonce}_{\mathrm{init}} \gets \text{random}(32)$\;
$a \gets \text{DeriveX25519Priv}(\mathrm{nonce}_{\mathrm{init}}, \text{``dh-init''})$\;
$R_0 \gets \KDFtext(\mathsf{CDI}, \ECDHtext(a, B), \text{``root-key''}, 32)$\;
\BlankLine
\Comment{Initialize chain key for current date}
$d_0 \gets \text{current\_date}()$\;
$ck_0 \gets \KDFtext(R_0, \text{``ck-init''} \,\|\, d_0, 32)$\;
\BlankLine
\textbf{store} $(R_0, K_{\mathrm{hash}}, a, ck_0, d_0)$\;
\end{algorithm}

\paratitle{Two Layer PII Protection}
\label{sec:design:protection}
The \sys{} architecture is designed to protect any data field that is identified as PII, whether by an automated scanner or, more robustly, by explicit developer annotation at the logging call site. For any such identified field, the client enforces in-situ protection through a two-layer design that separates pseudonymization from encryption. When a log message containing PII is emitted, \sys{} first pseudonymizes detected PII fields via keyed hashing, then encrypts the resulting tokens with daily-rotated ephemeral keys. This design ensures that even if encryption keys are compromised, plaintext PII is never recovered---only pseudonymized tokens suitable for correlation analysis.

\paratitle{Daily Key Ratcheting} 
To defend against multi-snapshot adversaries and provide forward secrecy, the client derives a fresh encryption key for each day using a ratcheting mechanism (Algorithm~\ref{alg:key-ratchet}). Each day, the chain key $ck_t$ is used to derive both the next chain key $ck_{t+1}$ and the daily message key $K_t$ via a one-way key derivation function. Once $ck_{t+1}$ is derived, the old chain key is deleted, preventing backward derivation and ensuring that compromise of the current state does not expose prior logs.

\paratitle{Hierarchical Ratchet for Post-Compromise Security} 
\sys{} employs a hierarchical ratcheting mechanism to protect against advanced adversaries who compromise a short-term chain key and attempt to recover all future logs. While daily ratcheting provides forward secrecy for past logs, it does not protect future logs if the current chain key is compromised. To counter this threat and preserve secrecy guarantees throughout the system's lifetime, \sys{} enforces a root key rotation whenever logs are exported to a server for analysis. Note in the sharing protocol described in \ref{alg:controlled-sharing}, providing the chain key $ck_{t^*}$ is an intentional compromise of the client's internal state. Viewed in a secure messaging context, all encrypted logs can be thought of as undelivered messages (as the client has not informed the server of an outer ratchet update). Once the client reveals any chain key via the controlled sharing protocol, it must rotate the root key $R_0$ by mixing in fresh entropy from the ephemeral Diffie--Hellman exchange performed during the sharing protocol. This ensures the new root key is unpredictable and severs the cryptographic chain from the old state, guaranteeing post-compromise security (G5).

\begin{algorithm}[t]
\caption{Daily key ratcheting}
\label{alg:key-ratchet}
\Input{Current chain key $ck_t$}
\Output{Next chain key $ck_{t+1}$ and message key $K_t$}
\SetKwFunction{FnKDFCK}{KDF\_CK}
\SetKwProg{Fn}{Function}{:}{\KwRet}
\Fn{\FnKDFCK{$ck$}}{
  $x \gets \KDFtext(ck, \text{``ratchet''}, 64)$\;
  $ck' \gets x[0{:}32]$\;
  $mk \gets x[32{:}64]$\;
  \KwRet $(ck', mk)$
}
\BlankLine
\Comment{Daily key derivation}
$(ck_{t+1}, mk_t) \gets \text{KDF\_CK}(ck_t)$\;
$mk_t$ \Comment*[r]{Field encryption key for day $t$}
\BlankLine
\textbf{store} $ck_{t+1}, mk_{t+1}$ \Comment*[r]{Update chain state}
\textbf{delete } $ck_t, mk_t$\;
\end{algorithm}

\paratitle{PII field protection during log emission} 
Algorithm~\ref{alg:pii-protection} details the two-layer protection applied to each detected PII field. In the first layer, the plaintext PII field $p$ is pseudonymized by computing a keyed HMAC using a server-specific hash key $K_{\mathrm{hash},t}$ derived from the long-term hash key $K_{\mathrm{hash}}$ (which is never shared). This produces a correlation token that enables linkage analysis without revealing the original value. In the second layer, the token is encrypted with the daily encryption key $K_t$ using authenticated encryption. The resulting ciphertext is stored in the log, ensuring that on-device observers, network adversaries, and multi-snapshot attackers cannot correlate PII across log dumps.

\begin{algorithm}[t]
\caption{PII field protection (log emission)}
\label{alg:pii-protection}
\Input{PII plaintext field $p$, $K_{\mathrm{hash}}$, encryption key $K_t$ for day $t$, server\_id, date $t$}
\Output{Protected field ciphertext $c$}
\Comment{Layer 1: Pseudonymization via keyed hashing}
$\mathrm{token} \gets \HMAC(K_{\mathrm{hash}}, p)$ \Comment*[r]{Hash plaintext PII}
\BlankLine
\Comment{Layer 2: Encryption with daily-rotated key}
$\mathrm{nonce} \gets \text{random}(12)$\;
$c \gets \AEAD.\Enc(mk_t, \mathrm{nonce}, \mathrm{token}, \emptyset)$\;
\BlankLine
\KwRet $c$ \Comment*[r]{Ciphertext stored in log}
\end{algorithm}

\paratitle{Controlled Sharing Protocol (Client Side)}
\label{sec:design:sharing:client}
To grant a forensic server time-windowed access to logs, the client initiates a controlled sharing protocol. The client exports the ratchet state for a specified interval, from a chosen start date $t^*$ to the current day, enabling the server to derive daily decryption keys for that period only. The client never shares the hash key $K_{\mathrm{hash}}$, ensuring that decryption yields only correlation tokens rather than plaintext PII. Immediately after export, the client rotates its root key $R_0$ (as described in \S\ref{sec:design:protection}), ensuring the server cannot decrypt logs generated after the current day. This provides post-compromise security and strictly bounds the decryption window. Algorithm~\ref{alg:controlled-sharing} presents the client-side protocol.

\begin{algorithm}[t]
\caption{Controlled sharing protocol}
\label{alg:controlled-sharing}
\small
\begin{minipage}[t]{0.47\textwidth}
\textbf{CLIENT SIDE:}
\BlankLine
\Input{Ephemeral server public key $B'$, start date $t^*$, current date $d$}
\Output{Grant sent to server}
\Comment{Generate ephemeral ECDH keypair}
$\mathrm{nonce}_{\mathrm{grant}} \gets \text{random}(32)$\;
$a' \gets \text{DeriveX25519Priv}(\mathrm{nonce}_{\mathrm{grant}}, \text{``export-keygen''})$\;
$A' \gets a' \cdot G$\;
\BlankLine
\Comment{Derive export encryption key}
$z \gets \ECDHtext(a', B')$\;
$K_{\mathrm{exp}} \gets \KDFtext(z, \text{``export-kdf''}, 32)$\;
\BlankLine
\Comment{Re-derive $ck_{t^{*}}$}
\If{$t^* < d$}{
$ck_0 \gets \KDFtext(R_0, \text{``ck-init''} \,\|\, d_0, 32)$\;
\For{$t \gets d_0$ \KwTo $t^{*} - 1$}{
  $(ck_{t+1}, \ \_) \gets \text{KDF\_CK}(ck_t)$ \Comment{$mk_t$ unused}
}
}
\Comment{Package and encrypt grant}
$\mathrm{AAD} \gets \mathrm{server\_id} \,\|\, \mathrm{device\_id} \,\|\, \mathrm{attest\_digest}\,\|\, \mathrm{grant\_id} \,\|\, d$\;
$M \gets \AEAD.\Enc(K_{\mathrm{exp}}, 
\{ck_{t^*}, t^*\}, \mathrm{AAD})$\;
\BlankLine
\textbf{send} $(A', M, \mathrm{AAD})$ to server\;
\BlankLine
\Comment{Post-grant rotation for forward secrecy}
$R_0' \gets \KDFtext(R_0, z, \text{``root-key-rotate''}, 32)$\;
$ck_0' \gets \KDFtext(R_0', \text{``ck-init''} \,\|\, (d{+}1), 32)$\;
\textbf{delete} $\{R_0, ck_{d}, a, d_0\}$\;
\textbf{store} $(R_0', ck_0', a', d)$\;
\end{minipage}
\par\noindent\rule{\linewidth}{0.4pt}\par
\begin{minipage}[t]{0.47\textwidth}
\textbf{SERVER SIDE:}
\BlankLine
\Input{Server ephemeral keypair $(b', B')$, grant date $d$, received $(A', M, \mathrm{nonce}, \mathrm{AAD})$ from client}
\Output{Daily keys $\{K_t\}$ for $t \in [t^*, d]$}
\Comment{Derive shared export key}
$z \gets \ECDHtext(b, A)$\;
$K_{\mathrm{exp}} \gets \KDFtext(z, \text{``export''}, 32)$\;
\BlankLine
\Comment{Decrypt and verify grant}
$\{ck_{t^*}, t^*\} \gets \AEAD.\Dec(K_{\mathrm{exp}}, \mathrm{nonce}, M, \mathrm{AAD})$\;
\textbf{if} verification fails \textbf{then} abort\;
\BlankLine
\Comment{Forward-only key derivation}
\For{$t \gets t^*$ \KwTo $d$}{
  $(ck_{t+1}, mk_t) \gets \text{KDF\_CK}(ck_t)$\;
  $K_t \gets mk_t[0{:}32]$\;
  \textbf{store} $K_t$ for day $t$\;
}
\end{minipage}
\end{algorithm}

\subsection{Server-Side Design}
\label{sec:design:server}

\paratitle{Controlled Sharing Protocol (Server Side)}
\label{sec:design:sharing:server}
The server receives the encrypted grant from the client and uses its long-term private key to derive the shared export key. After decrypting and verifying the grant, the server obtains the intermediate chain key $ck_{t^*}$ for the start date $t^*$. The server then employs a \emph{replay ratchet} mechanism: it initializes its local ratchet state with $ck_{t^*}$ and iteratively applies the same one-way key derivation function used by the client (Algorithm~\ref{alg:key-ratchet}) to derive daily decryption keys for each day in the authorized time window $[t^*, \text{current}]$. This forward-only replay ensures that the server can decrypt logs only within the granted interval, without the ability to derive keys for dates before $t^*$ or after the current day (due to the client's post-export root key rotation). The server-side protocol is shown in the right column of Algorithm~\ref{alg:controlled-sharing}.

\paratitle{Token Recovery and Forensic Analysis}
\label{sec:design:analysis}
After receiving daily decryption keys via the controlled sharing protocol, the forensic server decrypts protected log fields to recover correlation tokens. Algorithm~\ref{alg:token-recovery} details the token recovery process. For each ciphertext $c$ in the logs, the server applies the corresponding daily key $K_t$ to obtain the pseudonymized token. These tokens enable linkage analysis, timeline reconstruction, and cross-device correlation without exposing plaintext PII. Crucially, since the server never receives the hash key $K_{\mathrm{hash}}$, it cannot reverse the pseudonymization to recover the original sensitive values. The semantic structure of logs remains intact. The server can observe event sequences, construct attack graphs, and reconstruct timelines, while preserving privacy even under an honest-but-curious server model.

\begin{algorithm}[t]
\caption{Pseudonymized token recovery (server side)}
\label{alg:token-recovery}
\Input{Protected field ciphertext $c$ from log, decryption key $K_t$ for day $t$}
\Output{Pseudonymized token for PII correlation}
\Comment{Decrypt to recover pseudonymized token}
$\mathrm{token} \gets \AEAD.\Dec(K_t, \mathrm{nonce}, c, \emptyset)$\;
\textbf{if} decryption fails \textbf{then} abort\;
\BlankLine
\KwRet $\mathrm{token}$ \Comment*[r]{Used for correlation}
\end{algorithm}

We implemented \sys{} in two configurations: an end-to-end prototype in Python for evaluating the cryptographic protocols and macro-level performance on bulk log data, and an Android logging library for micro-benchmarking on real-world devices. The Android library provides an API that transparently integrates with existing applications, performing the client-side PII protection pipeline in user space. A detailed description of both implementations is provided in Appendix~\ref{sec:implementation}.

%% file: chapters/evaluation.tex
\section{Evaluation}
\label{sec:evaluation}

We evaluate \sys{} using two complementary approaches: an end-to-end prototype for macro-level measurements (storage overhead, bulk processing latency) and an Android library deployment for micro-level performance analysis (per-message latency, throughput). Both implementations are detailed in Appendix~\ref{sec:implementation}. Our evaluation addresses the following research questions:

\begin{enumerate}
    \item[RQ1] \textbf{Feasibility:} Can \sys{} operate within practical performance boundaries for production logging systems across diverse hardware generations?
    \item[RQ2] \textbf{Storage Efficiency:} What storage overhead does \sys{}'s two-layer protection introduce across different PII types, and how does it compare to plaintext logging?
    \item[RQ3] \textbf{Performance Scalability:} How do per-message latency and sustained throughput scale across devices with varying capabilities, and how does \sys{} compare to native Android logging?
    \item[RQ4] \textbf{Processing Bottlenecks:} Which pipeline components (key derivation, PII detection, encryption) dominate processing time, and where should optimization efforts focus?
    \item[RQ5] \textbf{Real-World Applicability:} Can \sys{} handle realistic log workloads at scale, processing log entries under high message rates with acceptable latency for both client and server operations?
    \item[RQ6] \textbf{Security Guarantees:} Does \sys{} provably achieve confidentiality of PII, forward secrecy against multi-snapshot adversaries, and integrity of exported logs under the threat model in \xref{sec:threatmodel}?
\end{enumerate}

\subsection{Experimental Setup}
\label{subsec:experimental-setup}

\paratitle{Test Devices}
Most modern smartphones ship with the requisite hardware support for DICE attestation. To demonstrate \sys{}'s feasibility and performance across a wide range of hardware, including legacy devices, we selected three generations of Android devices spanning 2017--2022. All three of our test devices (Table~\ref{tab:test-devices}) possess the necessary hardware capabilities to support a DICE implementation~\cite{android_key_attestation}. However, at the time of our evaluation, a stable, publicly available Android API that enables third-party applications to access the DICE framework was not readily available. This ecosystem limitation, rather than a hardware constraint, necessitated simulating the presence of a DICE provider to supply the initial hardware-rooted keys. This methodology allows us to precisely measure the performance overhead of the \sys{} cryptographic pipeline itself, which is the primary focus of our evaluation. Our results, therefore, demonstrate the practicality of the logging framework across a range of capable hardware, assuming that OS-level DICE APIs become available.

\paratitle{End-to-End Prototype}
For the end-to-end prototype evaluation, we use an Ubuntu 20.04 system with 32~GB RAM. The evaluations are not memory- or computationally intensive, so the prototype runs readily on any commodity PC. We plan to open-source our implementation for the broader research community.

\begin{table}[t]
\centering
\caption{Test devices used for stress testing \sys{} client performance. SD: SnapDragon, GB6-MC: Geekbench 6 multi-core scores~\cite{nanoreview_sd835_geekbench6,nanoreview_sd855_geekbench6,nanoreview_tensor_g1_geekbench6}.}
\label{tab:test-devices}
\footnotesize
\begin{tabular*}{\columnwidth}{@{\extracolsep{\fill}}lp{1.2cm}ccc@{}}
\toprule
\textbf{Device} & \textbf{SoC} & \textbf{RAM} & \textbf{GB6-MC} & \textbf{OS} \\
\midrule
Pixel 2 (2017) & SD 835 & 4GB & 1508 & Android 11 \\
Tab S6 (2019) & SD 855 & 8GB & 2834 & Android 12 \\
Pixel 6a (2022) & Tensor G1 & 6GB & 3208 & Android 13 \\
\bottomrule
\end{tabular*}
\end{table}

\subsection{Macro Evaluation}
\label{subsec:end-to-end-evaluation}

We evaluated the end-to-end \sys{} prototype (described in \xref{sec:implementation}) using the LogHub dataset~\cite{zhu2023loghub}, a large collection of system logs spanning diverse applications, servers, and devices. We focused on Android device logs comprising 25 files with 30.3~million log entries, totaling 3.37~GB of raw data. Using this dataset, we simulated a realistic log streaming, collection, and analysis pipeline to measure \sys{}'s macro-level performance characteristics.

\begin{figure*}[!t]
    \centering
    \begin{subfigure}[b]{0.32\textwidth}
        \includegraphics[width=\linewidth]{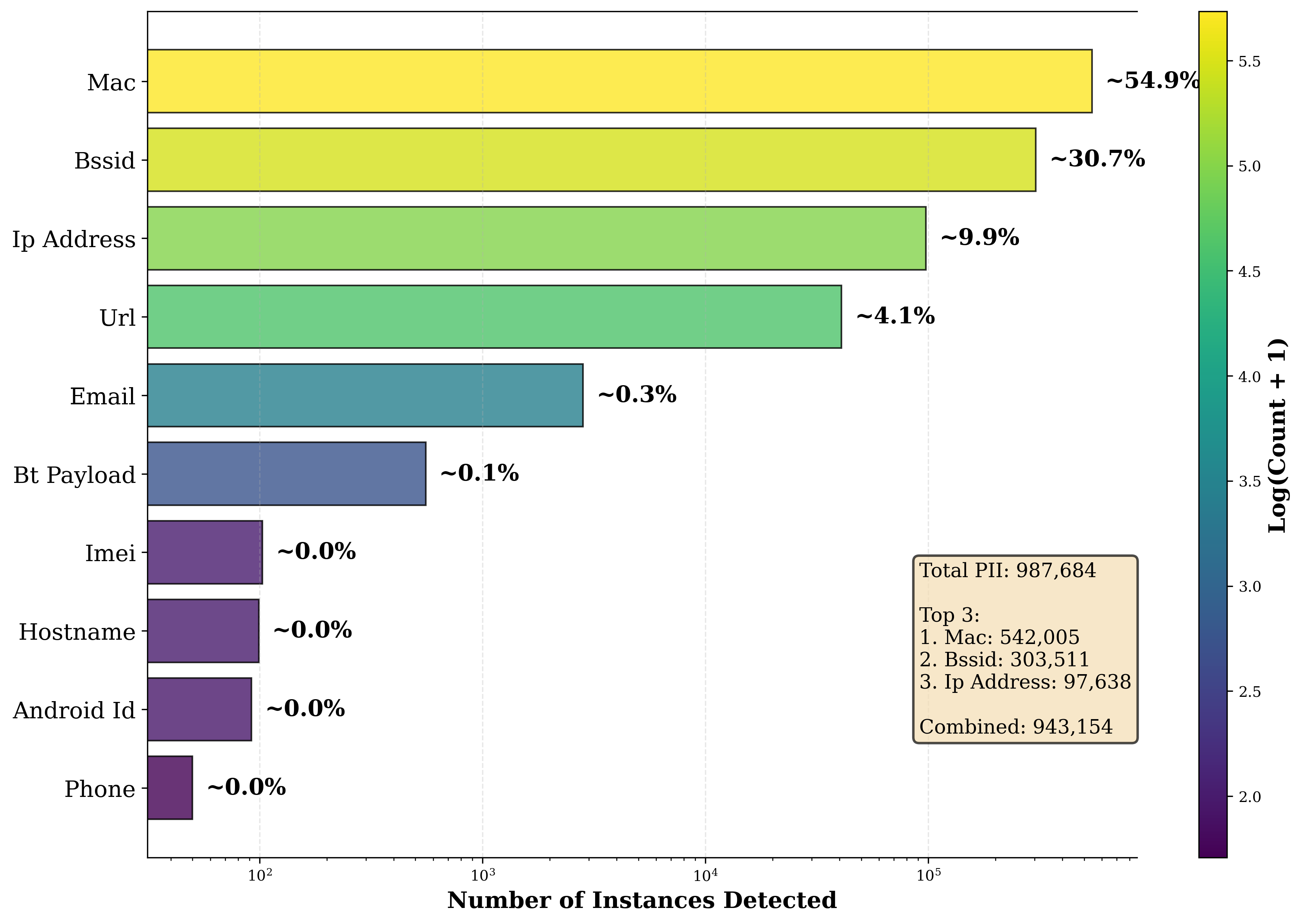}
        \caption{Distribution of PII occurrences in the dataset, as identified using standard Regex patterns.}
        \label{fig:pii-distribution}
    \end{subfigure}
    \hfill
    \begin{subfigure}[b]{0.32\textwidth}
        \includegraphics[width=\linewidth]{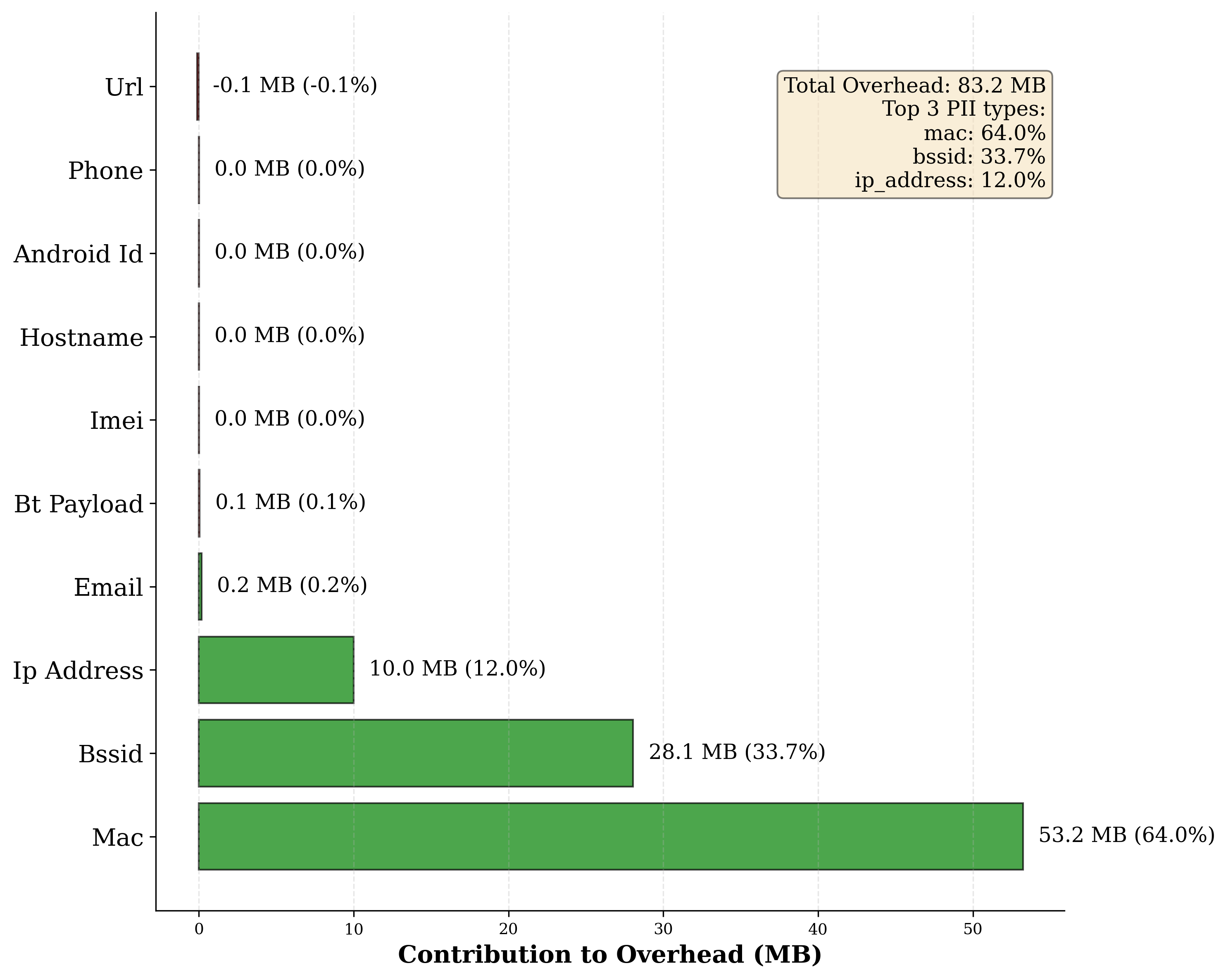}
        \caption{Relative contribution of each PII type introduced by \sys{} processing over the entire dataset.}
        \label{fig:pii-overhead-contributions}
    \end{subfigure}
    \hfill
    \begin{subfigure}[b]{0.32\textwidth}
        \includegraphics[width=\linewidth]{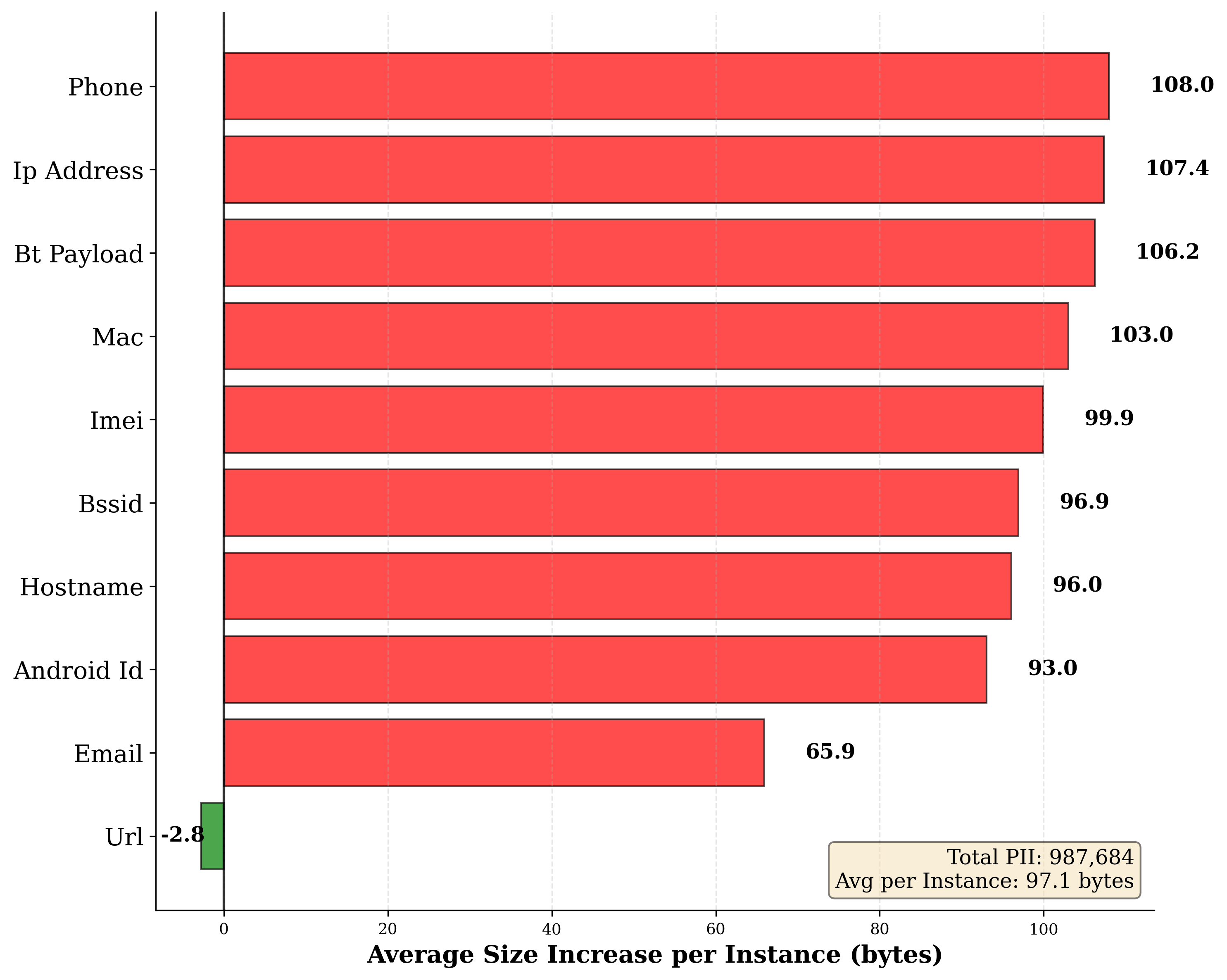}
        \caption{Storage overhead per PII instance. \sys{} reduces size for long PII types (\eg URLs with query parameters) due to fixed-size encrypted tokens.}
        \label{fig:size-overhead}
    \end{subfigure}
    \caption{PII analysis and storage overhead measurements using the end-to-end \sys{} prototype.}
    \label{fig:pii-analysis}
\end{figure*}

\paratitle{Size Overhead (RQ2)}
Figure~\ref{fig:pii-analysis} presents a granular breakdown of \sys{}'s storage efficiency across different PII types. Figure~\ref{fig:pii-distribution} shows the distribution of PII occurrences by type in the dataset. Passing the logs through \sys{}'s pipeline introduced 83.2~MB of additional overhead on top of the original 3.37~GB, resulting in a 2.41\% increase in size. Figure~\ref{fig:pii-overhead-contributions} shows the relative contribution of each PII type to this total overhead. Figure~\ref{fig:size-overhead} reveals an interesting property: \sys{} actually reduces storage for long PII types such as URLs. Many developers log complete URLs with query parameters (which often contain sensitive tokens) as single strings. Because \sys{} encrypts each URL as a fixed-size token, it reduces the per-occurrence overhead for URLs by 2.8~bytes on average compared to the original plaintext.

\begin{figure}[htbp]
    \centering
    \includegraphics[width=0.8\linewidth]{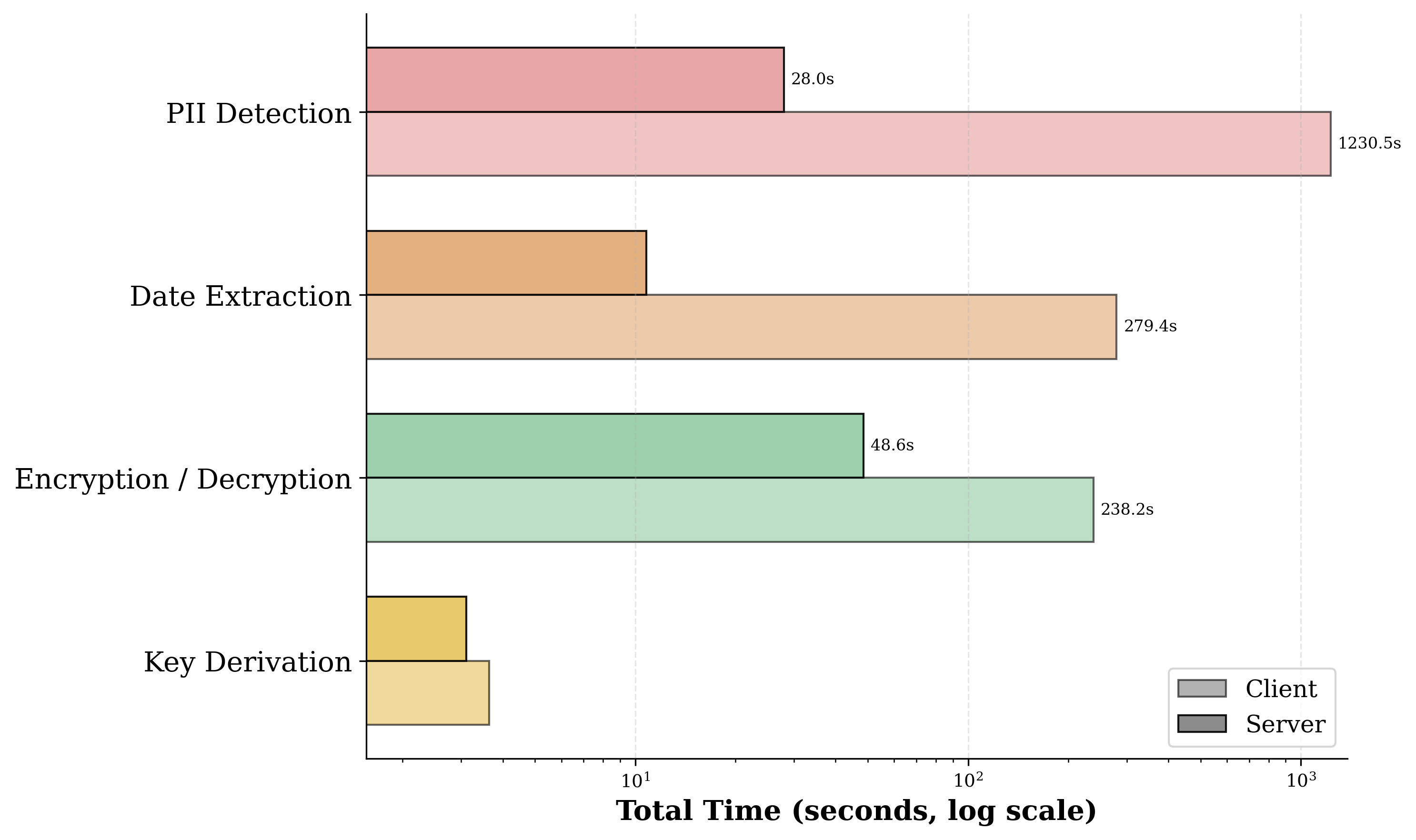}
    \caption{Latency budget observed by \sys{} client and server for each component during bulk processing of logs.}
    \label{fig:time-budget}
\end{figure}

\paratitle{Bulk Processing Latency (RQ4, RQ5)}
We evaluate the overall latency of \sys{}'s pipeline for bulk processing of the 30.3~million log entries in the dataset, addressing real-world applicability at scale (RQ5). Figure~\ref{fig:time-budget} breaks down the total time consumed by each component on both client and server sides. We categorize operations into four stages: (i)~PII Detection, (ii)~Date Extraction, (iii)~Encryption/Decryption, and (iv)~Key Derivation. The Date Extraction module parses timestamps from each log line to determine the appropriate daily key $K_t$ for encryption or decryption. PII Detection uses regex-based pattern matching to identify sensitive fields, and Encryption/Decryption follows the two-layer protection scheme described in \xref{sec:design:client}.

Client-side PII Detection step dominates the time budget (RQ4) because the client must apply all PII detection patterns (10 regex searches per log entry, corresponding to the PII types in Figure~\ref{fig:pii-distribution}) to identify sensitive fields. In contrast, the server only needs to locate pre-tagged \texttt{<PII>...</PII>} markers, which is significantly faster. Similarly, client-side encryption is slower than server-side decryption because of the additional overhead of in-place token replacement: after detecting PII, the client must locate and replace each plaintext occurrence with its encrypted token. When multiple PII fields appear in a single log line, this search-and-replace operation compounds. Despite these asymmetries, the total processing time remains practical for offline batch analysis of large log corpora.

\subsection{Micro Evaluation}
\label{subsec:on-device-evaluation}

\begin{figure*}[!t]
    \centering
    \begin{subfigure}[b]{0.48\textwidth}
        \includegraphics[width=\linewidth]{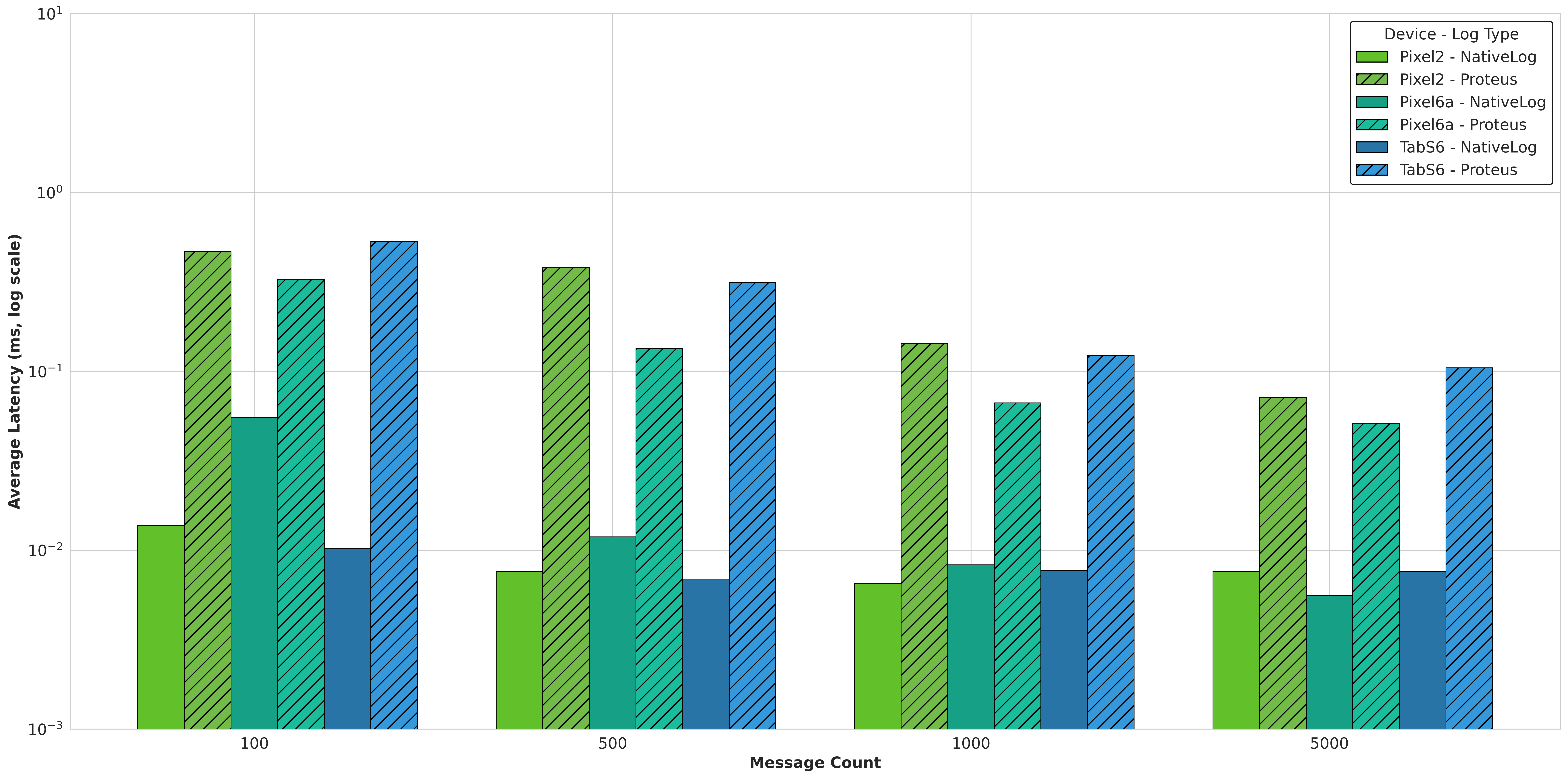}
        \caption{Per-message latency decreases with increasing message rate for \sys{}, while native Log latency stays constant.}
        \label{fig:latency-comparison}
    \end{subfigure}
    \hfill
    \begin{subfigure}[b]{0.48\textwidth}
        \includegraphics[width=\linewidth]{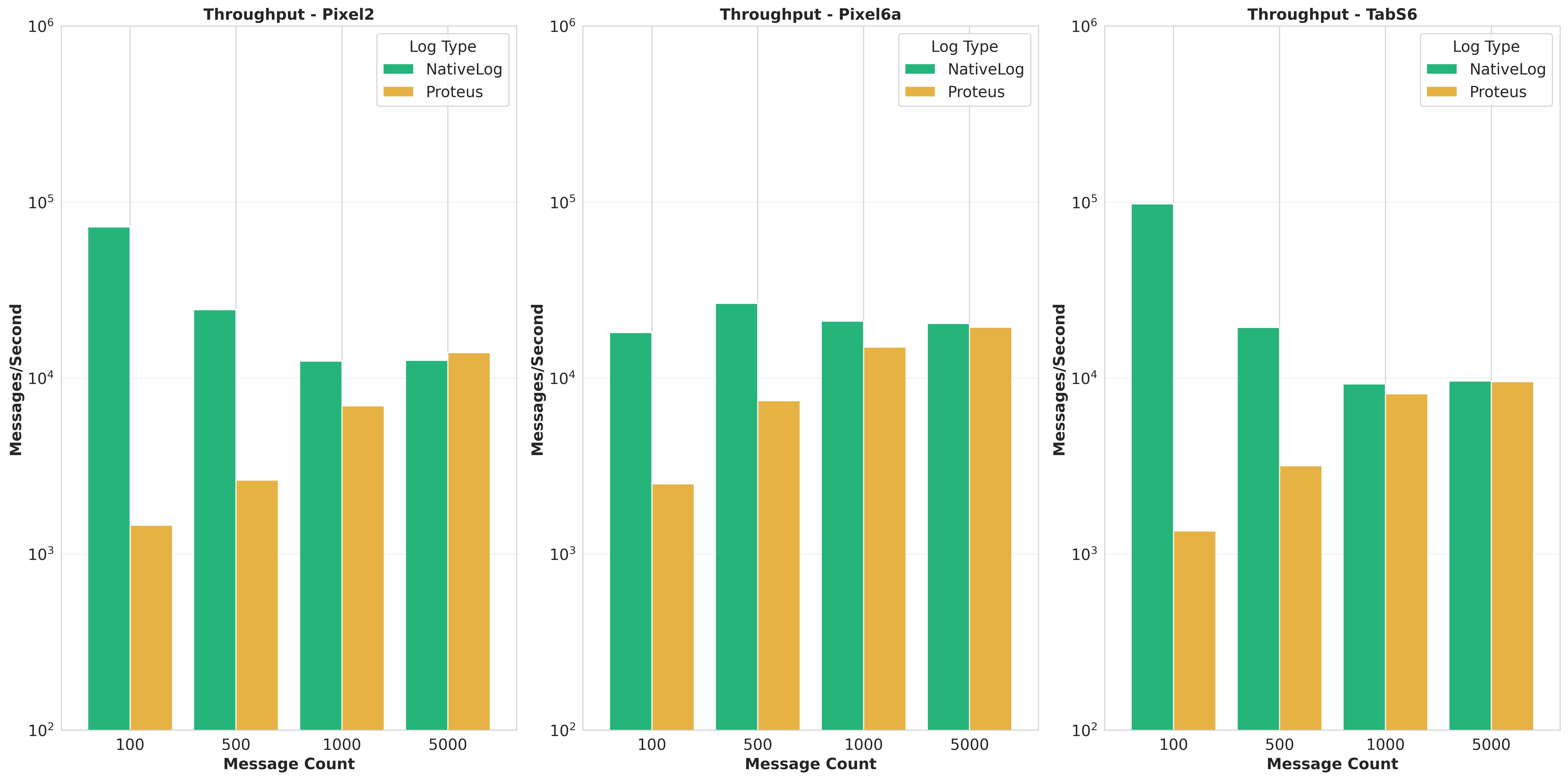}
        \caption{Observed throughput across devices under sustained workload shows that \sys{} can handle high message rates with equal throughput to native Log.}
        \label{fig:throughput-by-device}
    \end{subfigure}
    \\
    \begin{subfigure}[b]{0.48\textwidth}
        \includegraphics[width=\linewidth]{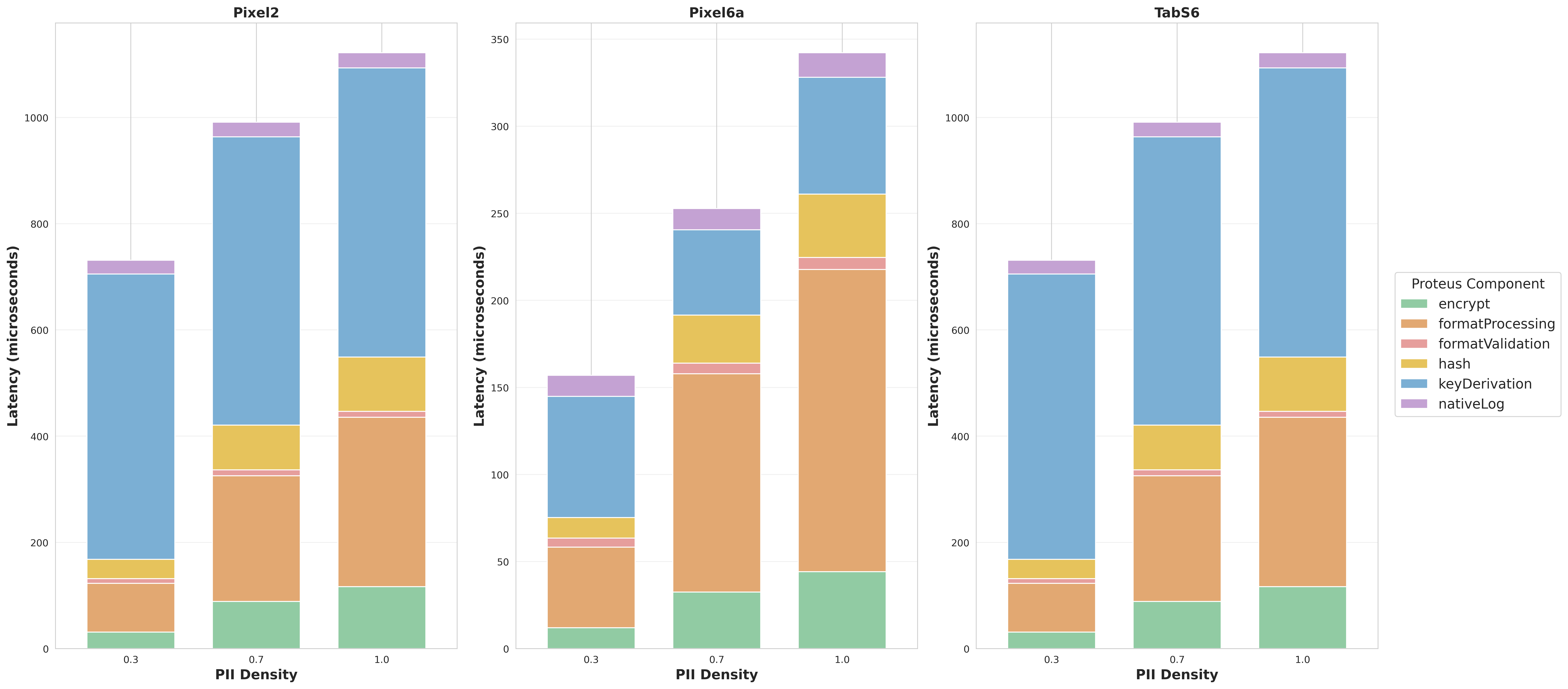}
        \caption{Component latency breakdowns across devices show that formatProcessing (Regex) and KeyDerivation (\sys{}) dominate the latency budget.}
        \label{fig:component-latency}
    \end{subfigure}
    \hfill
    \begin{subfigure}[b]{0.48\textwidth}
        \includegraphics[width=\linewidth]{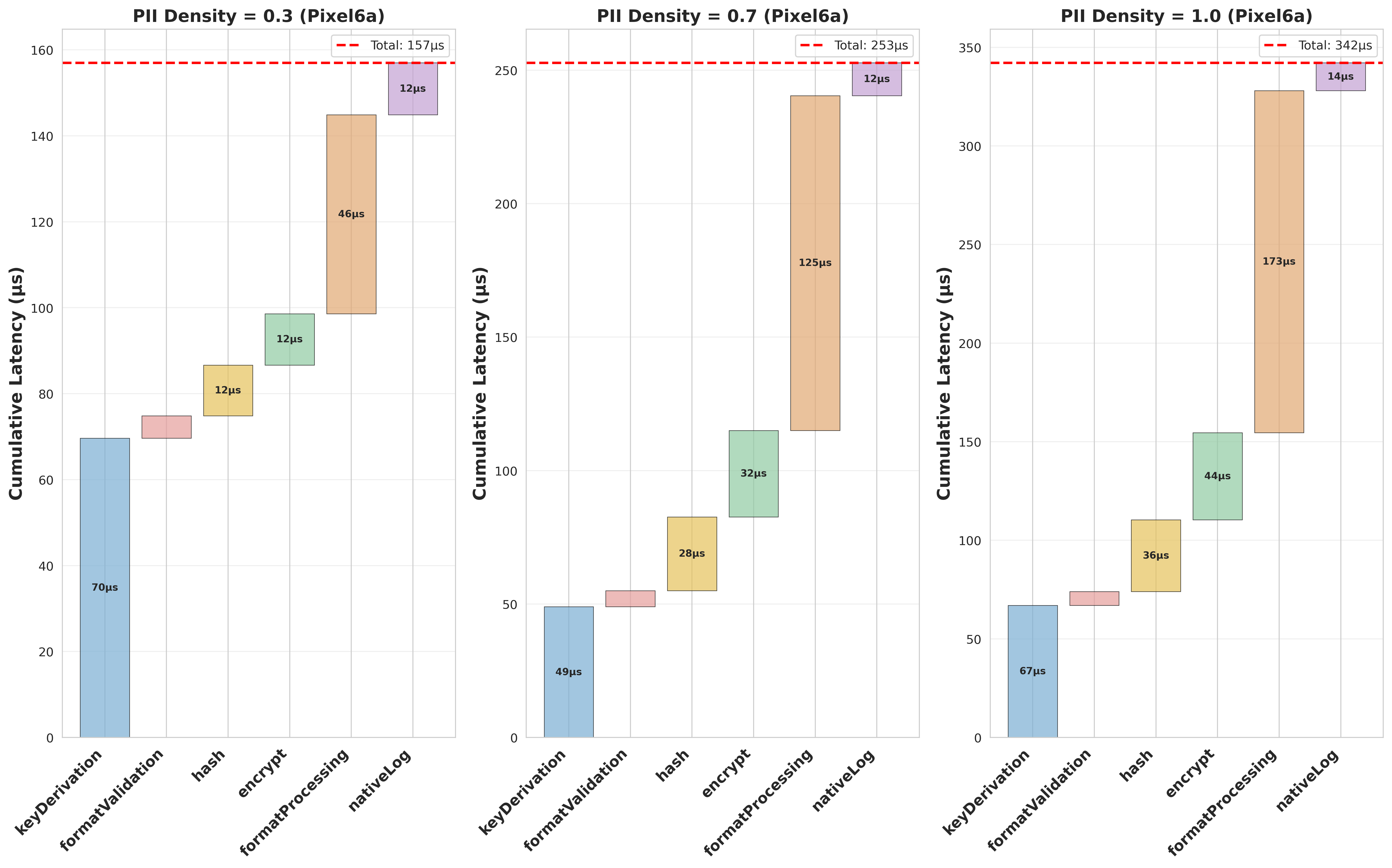}
        \caption{Varying the PII density per log line reveals FormatProcessing (Regex) as the actual bottleneck. Detection is supplementary to \sys{}(\xref{sec:discussion}).}
        \label{fig:waterfall-cumulative}
    \end{subfigure}
    \caption{Comprehensive on-device performance evaluation of \sys{} client.}
    \label{fig:on-device-comparison}
\end{figure*}

We deployed \sys{} as an Android library integrated with a sample application to conduct micro-level performance evaluation. We instrumented a stress-testing harness to exercise the logging pipeline under varying workloads, PII densities, and device capabilities. Figure~\ref{fig:on-device-comparison} presents the results across all three test devices.

\paratitle{Latency and Throughput (RQ3)}
Figure~\ref{fig:latency-comparison} and Figure~\ref{fig:throughput-by-device} present per-message latency and sustained throughput as a function of log rate, thereby assessing performance scalability across hardware generations. \sys{} incurs higher latency and lower throughput than the native Android \texttt{Log()} API, as expected, given that our implementation operates entirely in user space and performs additional cryptographic operations. Notably, latency decreases with increasing message rate, indicating that the overhead is likely dominated by context switching and garbage collection rather than cryptographic operations themselves. This suggests that a kernel-space implementation would significantly reduce per-message overhead.

\paratitle{Component-Level Breakdown (RQ4)}
Figure~\ref{fig:component-latency} shows the latency breakdown by component across all devices under varying PII densities (low, medium, high), identifying processing bottlenecks. Figure~\ref{fig:waterfall-cumulative} presents a waterfall view of the same breakdown on the Pixel 6a, the most performant device in our testbed. Two components dominate the latency budget: \texttt{keyDerivation} and \texttt{formatProcessing}. Key derivation computes the daily encryption key $K_t$ from the current date and chain key $ck_t$ for each logging operation. Format processing performs PII detection and in-place token replacement, which compounds when multiple PII fields appear in a single log line. Consistent with the bulk processing results in Figure~\ref{fig:time-budget}, PII detection and formatting consume the majority of processing time due to regex-based pattern matching and string manipulation. Optimization efforts should focus on these two stages: key derivation overhead could be amortized by caching daily keys or by implementing \sys{} within AOSP to perform key derivation once per day in kernel space.

\paratitle{Summary (RQ1, RQ5)}
Across our comprehensive evaluation suite, \sys{} demonstrates practical performance suitable for production logging systems. We establish feasibility by showing that \sys{} operates well within acceptable bounds for production systems: sub-millisecond per-message latency (0.2~ms median) and low single-digit storage overhead (2.41\%). These metrics are practical because they preserve the responsiveness of application logging (logging operations remain non-blocking) while incurring minimal storage costs that scale linearly with the number of PII occurrences, rather than with the total log volume (RQ1). The end-to-end prototype processed 30.3~million log entries, with bulk processing latency dominated by regex-based PII detection rather than cryptographic operations, demonstrating that the protection mechanisms are efficient. These results confirm that \sys{} can handle realistic log workloads at scale (RQ5), processing millions of log entries with acceptable latency for both client and server operations. While our user-space implementation incurs overhead from context switching and garbage collection, the results demonstrate that \sys{}'s two-layer protection scheme (\ie pseudonymization plus encryption with daily key ratcheting) operates within performant logging system boundaries. A kernel-space implementation would further reduce overhead by amortizing key derivation and eliminating user-space context switches, making \sys{} suitable for deployment in production logging frameworks. Security guarantees (RQ6) are formally established in \xref{sec:security-analysis}.

%% file: chapters/security_analysis.tex
\section{Security Analysis}
\label{sec:security-analysis}

Having presented \sys{}'s design (\xref{sec:design}) and empirical performance (\xref{sec:evaluation}), we now analyze its security properties, addressing RQ6. We establish that \sys{}'s hierarchical ratcheting construction provides provable confidentiality and forward secrecy by reducing its properties to Signal's messaging protocol, which has been formally analyzed under multi-snapshot adversaries. This section formalizes the security guarantees, identifies the cryptographic assumptions, and discusses limitations.

\subsection{Security Objectives}
\label{sec:security-analysis:goals}

We analyze \sys{} under the threat model defined in \xref{sec:threatmodel}: (i) a multi-snapshot on-device observer who captures repeated log dumps and internal client state, and (ii) an honest-but-curious server that receives exported logs and cryptographic grants. Building on the design in \xref{sec:design}, we establish three security objectives: (1) \emph{confidentiality of PII at rest and in transit}, ensuring that plaintext PII is never exposed in logs; (2) \emph{forward secrecy across time windows and access grants}, protecting past and future logs from compromise of current state; and (3) \emph{integrity and provenance of exported material}, binding logs to attested device identity.

The analysis assumes standard security notions for the underlying primitives: \KDFtext{} is a pseudorandom function family, \AEAD{} provides IND-CCA~\cite{bellare1998relations} confidentiality and integrity, and \HMAC{} is a keyed-pseudorandom function. We further assume that DICE attestation produces CDIs that are computationally indistinguishable from random, given device measurements~\cite{tao2021dice,tcg_dice_arch}.

\subsection{Comparison with Signal's Double Ratchet}
\label{sec:security-analysis:mapping}

The design in \xref{sec:design} employs a hierarchical ratcheting construction. We now show that it provides security guarantees analogous to Signal's well-analyzed double ratchet protocol~\cite{perrin2016doubleratchet,cohn2017signal,alwen2019double,bienstock2022more,collins2024tight}. \sys{} uses two nested ratchets: an \emph{outer ratchet} realized by the DICE-derived root key $R_0$, and an \emph{inner ratchet} realized by the daily chain key $ck_t$. The outer ratchet is explicitly rotated when an access grant is fulfilled (Algorithm~\ref{alg:controlled-sharing}, lines 13--15), and implicitly rotated whenever DICE re-measures the device firmware or boot state (since the CDI, and thus $R_0$, changes with different measurements). The inner ratchet advances daily via one-way key derivation (Algorithm~\ref{alg:key-ratchet}).

The key difference from Signal lies in the rotation triggers. Signal rotates its outer (Diffie--Hellman) ratchet whenever a party receives a fresh DH public key from its peer, starting a new sending/receiving chain with each message exchange. In contrast, \sys{} rotates its outer ratchet only on hardware-attested events (log export grants and firmware changes) and advances the inner ratchet on a deterministic daily schedule. We show that each epoch in \sys{} is functionally equivalent to a single (long) sending epoch in the double ratchet algorithm that omits the immediate decryption property. \sys{} thusly inherits the expected security properties analyzed in \cite{alwen2019double, bienstock2022more, collins2024tight}, See Appendix A.2.6 in \cite{bienstock2022more} for a thorough discussion on the immediate decryption property. In summary, it is a fundamental property for the practical usability of a secure messaging protocol. As such, immediate decryption is \emph{required} in their security model, but they acknowledge that it is not necessary to achieve both forward secrecy and post-compromise security.

\paratitle{Proof via Substitution}
We establish \sys{}'s security guarantees through a sequence of hybrid constructions that culminates in the initial sending epoch of the Signal double ratchet. Let $\mathcal{A}$ be a secure messaging adversary as formulated in \cite{alwen2019double, bienstock2022more, collins2024tight} attempting to distinguish any protected PII ciphertext in \sys{} from random. We define a sequence of hybrid constructions $\mathsf{H}_i$ that successively modify \sys{} that are indistinguishable to $\mathcal{A}$, but bring \sys{} closer to the original description of the double ratchet protocol~\cite{perrin2016doubleratchet}. We bound the corresponding change in the adversaries distinguishing advantage $\left|\Pr[\mathcal{A}(\mathsf{H}_i) \Rightarrow 1] - \Pr[\mathcal{A}(\mathsf{H}_{i+1}) \Rightarrow 1]\right|$ at each step.

\begin{itemize}
    \item \textbf{Hybrid $\mathsf{H}_0$ (Real \sys{} protocol).} This is the actual \sys{} deployment: CDI-rooted key derivation (Algorithm~\ref{alg:client-init}), two-layer PII protection (Algorithm~\ref{alg:pii-protection}), daily ratcheting (Algorithm~\ref{alg:key-ratchet}), and controlled sharing (Algorithm~\ref{alg:controlled-sharing}).

    \item \textbf{Hybrid $\mathsf{H}_1$ (Message reorganization).} Observe that each protected log for a given day is encrypted using the same daily message key $mk_t$. Thus, instead of inserting the output of algorithm \ref{alg:pii-protection} into a log entry, collect each ciphertext $c$ into a single stream $C$. Before running algorithm \ref{alg:key-ratchet} for the next day, emit $C$. This is simply a semantic change. An adversary who wishes to learn the tokens for each log entry will only be concerned with the emitted ciphertexts. In addition, each token is a fixed length, so there is no information lost from the view of an adversary's view of $C$ versus each individual $c$. Thus, $\mathsf{H}_0 = \mathsf{H}_1$

    \item \textbf{Hybrid $\mathsf{H}_2$ (Time dilation).} In this hybrid, we dispense with the notion of an (explicit) time for the rotation of the inner ratchet keys. Instead of emitting $C$ at the end of each day $t$, $C$ is emitted by the client before the next stream $C'$ is ready to be constructed. The counter $t$ will simply keep track of the number of sent messages, rather than an explicit timer for the number of days that have passed. Again, this is merely a notational change, and $\mathsf{H}_1 = \mathsf{H}_2$.

    \item \textbf{Hybrid $\mathsf{H}_3$ (Enabling Immediate Decryption).} In this hybrid, we assume that before the execution of \ref{alg:client-init}, the Dice CDI is shared and agreed-upon by the server. Here, the CDI now acts as a pre-shared key. This obviously allows for the server to decrypt each $C$ that it receives, but to an eavesdropping adversary, their view is identical to that in $\mathsf{H}_2$. Observe now that $\mathsf{H}_3$ is identical to the instantiation of a Forward-Secure AEAD as described in \cite{bienstock2022more} \S 4.2.3.

    \item \textbf{Hybrid $\mathsf{H}_4$ (Simplification).} In this hybrid, lines 6-13 in algorithm \ref{alg:controlled-sharing} are no longer executed. As in $\mathsf{H}_3$, the server was already able to compute chain keys $ck_t$, sending them is redundant. From the perspective of an eavesdropper, these keys were encrypted using a fresh ephemeral Diffie-Hellman exchange. In order to obtain them, they must be able to compute $z$ (line 5, Alg. \ref{alg:controlled-sharing}). Therefore $\left|\Pr[\mathcal{A}(\mathsf{H}_4) \Rightarrow 1] - \Pr[\mathcal{A}(\mathsf{H}_{3}) \Rightarrow 1]\right| \leq \epsilon_{DH}$, where $\epsilon_{DH}$ is $\mathcal{A}$'s advantage in computing $z$.
\end{itemize}

Notice that the protocol described in $\mathsf{H}_4$ is identical to that of the signal double ratchet \cite{perrin2016doubleratchet}, where the input to algorithm~\ref{alg:controlled-sharing} can be thought of as the client receiving a message from the server and progressing its outer-ratchet. The only significant difference from an adversary's point-of-view is the bundle of chain keys sent from the client to the server. Since every security proof of the double ratchet requires at least the Decisional Diffie-Hellman assumption, this is of no consequence. Since Signal's double ratchet provably provides confidentiality and break-in recovery under multi-snapshot compromise~\cite{cohn2017signal, alwen2019double, bienstock2022more, collins2024tight}, the final step of the proof shows that \sys{} achieves analogous security guarantees to a secure messaging adversary, with a daily message-sending epoch (inner ratchet) and post-grant root rotation (outer ratchet). The coarser rotation granularity compared to Signal means exposure windows are longer (up to one day for the inner ratchet), but the post-grant rotation in Algorithm~\ref{alg:controlled-sharing} ensures that any granted window has cryptographically independent keys from future logs, satisfying the security objectives in \xref{sec:threatmodel}.

\subsection{Security Guarantees}

\paratitle{Confidentiality of PII}
\label{sec:security-analysis:confidentiality}
Crucially, \sys{}'s two-layer design provides confidentiality even against the honest-but-curious server. The server never receives $K_{\mathrm{hash}}$ or the device CDI (only the ratchet state used to derive the decryption keys $mk_t$). Therefore, decryption of $c^\star$ reveals only the pseudonymized token $\text{token}^\star$ $= \HMAC(K_{\mathrm{hash}}, p_b)$, which is pseudorandom under the PRF assumption for \HMAC{}. Both the on-device adversary (who sees only encrypted ciphertexts without $K_t$) and the honest-but-curious server (who obtains $K_t$ but not $K_{\mathrm{hash}}$) observe only pseudorandom values. The only leakage is token equality (\ie identical PII values produce identical tokens), which is necessary for forensic correlation and explicitly permitted by design goal G3.

\paratitle{Forward Secrecy and Break-In Recovery}
\label{sec:security-analysis:forward-secrecy}
Forward secrecy follows directly from the ratchet construction. Consider an adversary who compromises the client at day $t$ and obtains the current state $(R_0, ck_t, K_t)$. Due to the one-way property of the inner ratchet (Algorithm~\ref{alg:key-ratchet}), recovering any prior chain key $ck_{t'}$ for $t' < t$ is computationally infeasible under the PRF assumption for \KDFtext{}. Thus, logs encrypted with keys $K_{t'}$ for $t' < t$ remain confidential even after state compromise at day $t$.

For post-compromise security (break-in recovery), \sys{} employs the hierarchical ratcheting mechanism described in \xref{sec:design:protection}. After each access grant, the client rotates the outer ratchet by re-deriving a fresh root key $R_0'$ from the current $R_0$ and the shared secret $z$ from the ephemeral DH exchange for that grant. This ensures that an adversary who has compromised the device state cannot predict future root keys, as they do not know the ephemeral private keys used in subsequent sharing protocols. This provides strong break-in recovery.

\paratitle{Integrity and Attestation Guarantees}
\label{sec:security-analysis:integrity}
The integrity of exported material follows from three mechanisms. First, the controlled sharing protocol (Algorithm~\ref{alg:controlled-sharing}) includes comprehensive context in the AEAD associated data: server identifier, device identifier, attestation digest, grant identifier, and grant date (lines 7--9). This binds the grant to the specific attested device state, server, and time window. Any tampering with the grant or its context is detected by AEAD verification (line 19). Second, since all symmetric keys are derived from the hardware-rooted CDI via \KDFtext{}, a server successfully decrypting a grant verifies that the client possesses the correct CDI and executed on a measured state authorized by the attestation policy. Third, each protected PII field carries an AEAD tag (Algorithm~\ref{alg:pii-protection}) that cannot be forged without knowledge of the daily key $mk_t$. Consequently, any alteration to protected log fields is detected during token recovery (Algorithm~\ref{alg:token-recovery}), and a malicious party cannot inject fabricated PII tokens that appear authentic.

\paratitle{Limitations}
\label{sec:security-analysis:limitations}
The analysis inherits the standard assumptions of the underlying cryptographic primitives and security proofs~\cite{cohn2017signal}. We identify three limitations. First, if an adversary compromises $K_{\mathrm{hash}}$ (which remains on-device for the lifetime of the CDI), they can compute $\HMAC(K_{\mathrm{hash}}, \cdot)$ for candidate PII values and link tokens to plaintext through offline dictionary attacks. This is mitigated by the hardware binding of $K_{\mathrm{hash}}$ to the CDI, requiring physical device access and extraction. Second, \sys{}'s coarser rotation granularity (daily for inner ratchet, per-grant for outer ratchet) creates longer exposure windows than Signal's per-message rotation. A compromise at day $t$ exposes all logs encrypted with $K_t$ on that day. However, this granularity is appropriate for logging workloads, where sub-second key rotation would impose prohibitive overhead. Third, metadata leakage, including timestamps, log categories, message sizes, and ciphertext lengths, remains observable to all parties and can reveal behavioral patterns. As stated in \xref{sec:threatmodel}, metadata privacy and traffic analysis resistance are explicitly out of scope for \sys{}.

%% file: chapters/discussion.tex
\section{Discussion}
\label{sec:discussion}

\paratitle{PII Detection Strategy}
Our prototype implements regex-based PII detection as a stand-in for other production-grade inference-based identification mechanisms~\cite{microsoft_presidio,mainetti2025detecting,savkin2025spy}. While this is sufficient to evaluate the performance of our cryptographic pipeline, we do not claim it is a complete solution to the PII detection problem. Indeed, our work highlights that the most secure and performant path forward is to shift responsibility from brittle, post hoc scanning to the developer. We advocate for the adoption of privacy-aware logging APIs, where sensitive data is explicitly tagged at the source (\eg \texttt{Log.safe("Login for \%s", pii(user\_email))}). \sys{} provides the ideal backend architecture to support such an API, guaranteeing protection once a field is marked, thereby solving the enforcement problem while leaving the (orthogonal) identification problem to a more appropriate layer.

\paratitle{DICE Attestation Practicality}
Our evaluation focuses on the performance of the \sys{} cryptographic pipeline, which operates on keys \emph{provided by} an attestation framework. By simulating this interface, we accurately benchmark the overhead of our contribution. While the overall end-to-end security depends on the hardware's correct implementation of DICE, our work demonstrates that the software layer built on it is practical and efficient.

\paratitle{Server-Side Scalability}
A potential concern is the complexity of managing decryption keys and ratchet states for thousands or millions of devices on the server side. However, the threat model for forensic analysis clarifies the scope of this challenge. PII token decryption is not intended for bulk, real-time analytics across all devices. Instead, it is a targeted forensic tool used on a small handful of devices that have already been identified as compromised or suspicious through other means. The initial stages of an investigation, such as timeline reconstruction and provenance analysis, can proceed on the encrypted logs without requiring PII correlation. The resource-intensive decryption and correlation process is only invoked selectively, ensuring the server-side workload remains manageable.

%% file: chapters/related_work.tex
\section{Related Work}
\label{sec:related}

\paratitle{Tamper-Evident and Secure Logging}
A significant body of work focuses on log integrity, ensuring logs cannot be altered post-facto. These systems provide tamper-evidence through techniques like cryptographic chaining~\cite{schneier1999secure}, Trusted Execution Environments (TEEs)~\cite{karande2017sgx,paccagnella2020custos}, or eBPF~\cite{zhao2025rethinking}. While essential for establishing log immutability, they protect the container rather than its contents. \sys{} provides a complementary guarantee of log \emph{provenance} via DICE attestation, binding logs to a device's state at creation. By protecting both the content and its origin, \sys{} offers an integrated solution for provenance, integrity, and confidentiality.

\paratitle{Log Reduction}
A parallel line of research focuses on log reduction and compression to manage high data volumes~\cite{lee2013loggc, tang2018nodemerge, sun2024auditrim, inam2022faust}. These systems use techniques like heuristic pruning and in-kernel filtering to reduce storage overhead. However, this provides data minimization, not content protection; any remaining log entries still contain plaintext PII. \sys{}'s in-situ protection is therefore complementary, ensuring confidentiality even within a reduced log stream.

\paratitle{Privacy-Preserving Data Collection}
Prior work on privacy preserving data collection fails to reconcile utility and confidentiality. Post-hoc redaction~\cite{microsoft_presidio} exposes PII during collection and transit. Client-side taint analysis~\cite{Enck2010TaintDroid,Arzt2014FlowDroid,hu2021samldroid,yang2022fsaflow} incurs high overhead and has coverage gaps. Differential privacy~\cite{erlingsson2014rappor,ren2022ldp,mannhardt2019privacy,varma2022sarve} destroys the event-level fidelity required for forensics. Even selective decryption of encrypted logs~\cite{chaulagain2024fa} still exposes plaintext PII. \sys{} resolves this trade-off by separating correlation from disclosure. It enables linkage analysis on pseudonymized tokens without ever exposing the sensitive plaintext values, providing an essential privacy layer for modern forensic systems~\cite{cheng2024kairos,zeng2022shadewatcher,jia2024magic,rehman2024flash,sekar2024eaudit}.

%% file: chapters/conclusion.tex
\section{Conclusion}
\label{sec:conclusion}

This paper presents \sys{}, the first in-situ logging framework that prevents PII disclosure while preserving full forensic utility for mobile endpoints. By exploiting the insight that forensic analysis requires correlation rather than the recovery of PII, \sys{} addresses longstanding challenges in privacy-preserving logging: protecting data at emission, defending against multi-snapshot adversaries, and enabling controlled server-side correlation without exposing plaintext. Our hardware-rooted double-ratchet construction with DICE-anchored key derivation provides provable forward secrecy, while the controlled-sharing protocol enables time-windowed forensic access to pseudonymized tokens. Evaluation across three Android device generations demonstrates practical overhead (median 0.2~ms latency, 2.41\% storage) suitable for production deployment, showing that privacy-preserving logging is achievable even under continuous exfiltration to honest-but-curious cloud platforms.

%% file: chapters/implementation.tex
\section{Implementation}
\label{sec:implementation}

We implemented \sys{} in two configurations to enable comprehensive evaluation of both cryptographic correctness and real-world performance.

\paratitle{End-to-End Prototype}
Our first implementation is an end-to-end prototype (client and server) in Python that evaluates the complete cryptographic primitives and logging utility construction. We used the LogHub dataset~\cite{zhu2023loghub}, the largest publicly available collection of system logs at the time of writing, to simulate a realistic log streaming, collection, and analysis pipeline built for Android devices. Figure~\ref{fig:component-diagram} shows the end-to-end architecture of this prototype.

The client component simulates DICE to derive a CDI and device-bound identity, from which it derives the three long-term keys ($R_0$, $K_{\mathrm{hash}}$, $s_{\mathrm{DH}}$) as specified in Algorithm~\ref{alg:client-init}. The client then issues an Ed25519-signed X.509 certificate embedding its X25519 public key and performs Diffie--Hellman-based key agreement to establish a shared export key $K_{\mathrm{exp}}$. Log protection operates as a streaming pipeline: per-line date extraction, regex-based PII detection, per-day symmetric ratcheting (HKDF) to derive daily encryption keys $K_t$ from the chain key $ck_t$, HMAC-based pseudonym tokenization using $K_{\mathrm{hash}}$, and AEAD encryption (Fernet) with $K_t$. The client emits tagged protected lines, records timing and space metrics, and exports time-bounded capabilities $(t^*, ck_{t^*})$ using the controlled sharing protocol for server-side replay. The server component loads a pinned client certificate, reconstructs the export key $K_{\mathrm{exp}}$ via Diffie--Hellman, and replays the symmetric ratchet forward from the provided chain key $ck_{t^*}$ to derive day-specific decryption keys $\{K_t\}$. For each protected log line, the server locates PII tags, derives the correct per-day key from the ratchet, decrypts to recover stable pseudonym tokens, and collects data for analysis.

The purpose of this prototype is to evaluate macro-level measurements using real-world logs: overall size increase, per-PII-type overhead, and end-to-end pipeline performance. These measurements, presented in \xref{sec:evaluation}, provide insight into the storage and bandwidth costs of \sys{} across diverse log workloads.

\paratitle{Android Logging Library}
Our Android implementation focuses on \sys{}'s client-side features for micro-evaluation of performance on real-world devices. We deployed the prototype across three hardware generations with varying capabilities: Pixel 2 (2017), Pixel 6a (2022), and Samsung Galaxy Tab S6 (2019). To ensure modularity and ease of integration with any Android application, we designed a user-space library \texttt{loglib} that can be imported into any app to leverage \sys{}'s logging framework. \sys{} as a logging library introduces a \texttt{Log.safe()} API that mimics the functionality of the native \texttt{Log()} API in AOSP. \texttt{Log.safe()} implements all the client-side functionality described in \xref{sec:design:client} using standard Android and Java cryptographic imports: DICE-based key derivation to obtain $R_0$, $K_{\mathrm{hash}}$, and $s_{\mathrm{DH}}$; regex-based PII detection; HMAC-based pseudonym tokenization with $K_{\mathrm{hash}}$; per-day symmetric ratcheting (HKDF) to derive daily encryption keys $K_t$ from chain key $ck_t$; and AEAD encryption with $K_t$.

The prototype instruments timing metrics for each pipeline stage (key derivation, pattern scan, hashing, encryption) at nanosecond granularity to enable fine-grained performance analysis. We include a stress-testing harness that exercises throughput, concurrency, latency, memory, and PII-density scenarios, with on-device result collection to app-scoped storage. We rely on conservative implementation patterns, avoiding reflection or native code paths in the core pipeline, to ensure reproducibility across devices.

Due to development constraints---specifically, limitations in modifying AOSP where the native \texttt{Log()} API is implemented---our solution is entirely implemented in user space. This design limitation introduces additional overhead in our latency experiments: garbage collection and context switching by user-space applications increase the measured latency compared to the AOSP-native \texttt{Log()} API. Despite these constraints, our measurements demonstrate that \sys{} imposes practical overhead suitable for production deployment. We discuss the performance results in detail in \xref{sec:evaluation}.